\documentclass[aps,pra,reprint,superscriptaddress,amsmath,amssymb,preprintnumbers,showpacs]{revtex4-1}
\usepackage[normalem]{ulem}

\usepackage{enumerate}
\usepackage{graphicx}
\usepackage{color}
\usepackage{float} 
\usepackage{xspace}
\usepackage{bm} 
\graphicspath{{./}}

\newcommand{\ket}[1]{\bigl| #1 \bigr>} 
\newcommand{\bra}[1]{\bigl< #1 \bigr|} 
\newcommand{\abs}[1]{\left| #1 \right|} 
\newcommand{\norm}[1]{\left\lVert#1\right\rVert}

\newcommand{\floqel}[1]{\xi_{E_\text{e}}}
\newcommand{\floqnuc}[1]{\xi_{E_\text{N}}}

\newcommand{\uv}[1]{\ensuremath{\mathbf{\hat{#1}}}} 
\renewcommand{\vec}[1]{{\mathbf{\bm{#1}}}}

\newcommand{\op}[1]{\hat{#1}}
\newcommand{\vop}[1]{\op{\vec{#1}}}

\newcommand{\hBN}{hBN}
\newcommand{\ehPE}{$eh$-PER}
\newcommand{\berryc}{\vec{\mathcal{A}}}

\newcommand{\com}{\vec{\mathcal{D}}}
\newcommand{\ccom}{\vec{\mathcal{Q}}}

\newcommand{\pol}{\mu}

\newcommand{\ellip}{\epsilon}
\newcommand{\hh}{H} 

\begin{document}
\title{Expanded view of electron-hole recollisions in solid-state high-harmonic generation: Significance of full-Brillouin-zone tunneling and imperfect recollisions}
\author{Lun \surname{Yue}}
\email{lun\_yue@msn.com}
\author{Mette B. \surname{Gaarde}}
\email{mgaarde1@lsu.edu}
\affiliation{Department of Physics and Astronomy, Louisiana State University, Baton Rouge, Louisiana 70803-4001, USA}
\date{\today}

\begin{abstract}
  We theoretically investigate electron-hole recollisions in high-harmonic generation (HHG) in band-gap solids irradiated by linearly and elliptically polarized drivers. We find that in many cases the emitted harmonics do not originate in electron-hole pairs created at the minimum band gap, where the tunneling probability is maximized, but rather in pairs created across an extended region of the Brillouin zone (BZ). In these situations, the analogy to gas-phase HHG in terms of the short- and long-trajectory categorizations is inadequate. Our analysis methodology comprises three complementary levels of theory: the numerical solutions to the semiconductor Bloch equations, an extended semiclassical recollision model, and a quantum wave packet approach. We apply this methodology to two general material types with representative band structures: a bulk system and a hexagonal monolayer system. In the bulk, the interband harmonics generated using elliptically-polarized drivers are found to originate not from tunneling at the minimum band gap $\Gamma$, but from regions away from it. In the monolayer system driven by linearly-polarized pulses, tunneling regions near different symmetry points in the BZ lead to distinct harmonic energies and emission profiles. We show that the imperfect recollisions, where an electron-hole pair recollide while being spatially separated, are important in both bulk and monolayer materials. The excellent agreement between our three levels of theory highlights and characterizes the complexity behind the HHG emission dynamics in solids, and expands on the notion of interband HHG as always originating in trajectories tunnelled at the minimum band gap. Our work furthers the fundamental understanding of HHG in periodic systems and will benefit the future design of experiments.
\end{abstract}

\maketitle


\section{Introduction} \label{sec:intro}

The recent experimental observations of high-harmonic generation (HHG) in solids \cite{Ghimire2011, Vampa2015Nat, You2017, Ndabashimiye2016, Garg2016, Wang2017} have contributed to the rapid progress of attosecond physics in the condensed matter phase \cite{Ghimire2014rev, Kruchinin2018rev, Li2020rev}. Solid-state HHG carries exciting prospects for the engineering of compact attosecond light sources \cite{Luu2015, Sivis2017, Han2016, Vampa2017, Gholam-Mirzaei2017, Garg2018, Yang2019} and ultrafast spectroscopy methods capable of probing band structures \cite{Vampa2015PRL, Uzan2020}, impurities \cite{Huang2017, Almalki2018, Yu2019, Chinzei2020}, and topological features \cite{Liu2017, Luu2018, Bauer2018, Silva2019, Chacon2020, Juerss2020, Bai2020, Baykusheva2021PRA}. The understanding of the HHG process has been aided by accurate computational quantum theories such as time-dependent density functional theory (TD-DFT) \cite{Runge1984, Tancogne-Dejean2017PRL, Yu2020} and semiconductor Bloch equations (SBEs) \cite{Golde2008, Kira2012}. These theories have helped to establish that the high harmonics with energies less than the band gap have large contributions from the \textit{intraband} currents originating in the electron-hole motion in their respective bands, while above-band-gap harmonics are generally dominated by the \textit{interband} currents originating in the coupling between the bands. While the SBEs and TD-DFT methods are able to accurately simulate the HHG process, they can be regarded as numerical experiments that contain all the relevant physics, and the underlying physical pictures can be difficult to extract. For this reason the celebrated gas-phase three-step model \cite{Corkum1993, Lewenstein1994} has been generalized to solids \cite{Vampa2014, Vampa2015PRB, Parks2020}, and has been shown to provide an intuitive real-space picture for the interband harmonics: an electron-hole pair is created when the external field causes an electron to tunnel from the valence band to the conduction band at the minimum band gap; the electron and hole are driven apart by the laser; they can recollide when they spatially reencounter each other, leading to the emission of harmonic radiation. The recollision picture has been instrumental in our fundamental understanding of solid-state HHG \cite{Vampa2014, Vampa2015PRB, Vampa2015Nat, McDonald2015, Zhang2019, Uzan2020}, as well as other related nonlinear phenomena such as high-order-sideband generation \cite{Liu2007AIPC, Zaks2012, Langer2016, Banks2017, Langer2018}.

The tunneling, propagation and recollision dynamics responsible for HHG in solids differ significantly from their counterpart in the gas phase. In gases, the continuum-electron is free and its dispersion is quadratic such that its group velocity is always along the direction of the canonical momentum. In crystalline solids, however, the quadratic dispersion only holds near certain high-symmetry points in reciprocal space. Consequently, the group velocities of the electron and hole are generally much more complex, and can even lead to imperfect recollisions where an electron-hole pair recollide even though their centers are spatially separated \cite{Crosse2014NC, Zhang2019, Yue2020PRL}. The exponential dependence of the tunneling rate on the band gap \cite{Keldysh1964} dictates that tunneling occurs with the highest probability at the minimum band gap. However, due to the complicated dispersions, electron-hole trajectories that originate away from the minimum band gap could have higher chances of recollision and end up dominating the emission process. Similarly, electron-hole pairs created near different symmetry points in the Brillouin zone (BZ) could lead to drastically different harmonic energies and emission time profiles. The full understanding of these complexities for HHG in solids are critical for the probing of the full BZ, as well as the design of new ultrafast light sources. The original semiclassical recollision model in solids, however, assumes tunneling at the minimum band gap and with perfect recollisions, and cannot provide a full framework for the HHG process in solids apart from simple cases. Due to these limitations, for example, the authors in \cite{Li2019} concluded that the recollision picture would fail for solid-state HHG with circularly polarized fields.

In this manuscript, we conclusively show that in many common experimental scenarios, and for several types of materials, the electron-hole pairs created away from the minimum band gap not only contribute to, but can strongly dominate the recollision-driven harmonic emission. In these cases, the understanding of the emission dynamics in terms of short and long trajectories, well-known from gas-phase HHG, breaks down. This breakdown can be due to either the band structure or the laser polarization, and we provide two examples of current experimental and theoretical interest: HHG in a generic bulk crystal induced by elliptically polarized fields, and HHG induced in a generic monolayer material by linearly polarized fields. Our analysis comprises three complementary levels of theory: the full numerical solution of the SBEs, an extended recollision model, and a model based on construction of recolliding electron-hole wave packets. We find that the novel harmonic emission profiles are due to the collective emission associated with trajectories originating in extended regions near different symmetry points. We show that tunneling from different regions can lead to different time-frequency emission characteristics, with impact for HHG-based optical probing of the whole BZ. The familiar short and long trajectories can be recovered in special cases: the bulk irradiated by linearly polarized field, and the monolayer irradiated by fields polarized along specific symmetry directions. We provide general rules for when one can expect the underlying physics in solid-state HHG to substantially deviate from that of the gas phase.

This paper is organized as follows. Section~\ref{sec:theory} contains the theoretical framework pertinent to this work: The SBEs are given in Sec.~\ref{sec:theory_sbe}, the semiclassical picture is detailed in Sec.~\ref{sec:theory_semiclass}, and the electron-hole wave packet construction is described in Sec.~\ref{sec:theory_wp}. Section~\ref{sec:zno2d} treats the ellipticity dependence of HHG in the representative bulk system zinc oxide (ZnO): the model is defined in Sec.~\ref{sec:zno2d_model}, the full quantum result from the SBEs are presented in Sec.~\ref{sec:zno2d_hhg}, the semiclassical analysis with tunneling at $\Gamma$ is discussed in Sec.~\ref{sec:zno2d_gamma}, and the full-BZ recollision picture is given in Sec.~\ref{sec:zno2d_full}. Section~\ref{sec:hbn2d} investigates the HHG in the representative monolayer system hexagonal boron nitride (\hBN): The model is described in Sec.~\ref{sec:hbn2d_model}, the wavelength dependence is explored in Sec.~\ref{sec:hbn2d_hhg}, the orientation dependence in Sec.~\ref{sec:hbn2d_orientation}, and the quantum wave packet analysis in Sec.~\ref{sec:hbn2d_wp}. Section~\ref{sec:hbn2d_conclusion} concludes the paper and provides outlook. Appendix~\ref{sec:app_derivs} provides details on the derivations of the saddle points equations, and Appendix~\ref{sec:app_suppl} includes relevant supplemental figures.
Atomic units are used throughout this work unless indicated otherwise.


\section{Theoretical methods} \label{sec:theory}

In this section, we describe the theoretical framework pertinent to this work. The numerical solutions to the SBEs can be considered a numerical experiment and is our ``full quantum'' benchmark result, which the semiclassical recollision and wave packet methods will be compared to. In the following subsections, we assume that relevant quantities such as the band structures, transition dipole moments, Berry connections and Berry curvatures are known in advance, either by employing model systems or using commercial solid-state structure codes \cite{Wien2k, Kresse1996}.

\subsection{Semiconductor Bloch equations} \label{sec:theory_sbe}
The SBEs governing a solid driven by a strong laser reads \cite{Golde2008, Kira2012, Schubert2014, Vampa2014, Jiang2018, Floss2018}
\begin{equation}
  \label{eq:theory_sbe_1}
  \begin{aligned}
    \dot{\rho}_{mn}^{\vec{K}}(t)
    =& -i\left[E_m^{\vec{K}+\vec{A}(t)}-E_n^{\vec{K}+\vec{A}(t)} - \frac{i(1 - \delta_{mn})}{T_2} \right]\rho_{mn}^{\vec{K}}(t)
    \\
    &- i\vec{F}(t)\cdot \sum_l
    \left[\vec{d}_{ml}^{\vec{K}+\vec{A}(t)} \rho_{ln}^{\vec{K}}(t) 
      - \vec{d}_{ln}^{\vec{K}+\vec{A}(t)} \rho_{ml}^{\vec{K}}(t)  \right],
  \end{aligned}
\end{equation}
with $\vec{K}$ the crystal momenta in a reciprocal reference frame moving with $\vec{A}(t)\equiv -\int^t\vec{F}(t')dt'$, $\vec{F}(t)$ the electric field, $E_n^{\vec{k}}$ the band energies, $\vec{d}_{mn}^{\vec{k}}=i\bra{u_m^{k}}\nabla_{\vec{k}}\ket{u_n^{\vec{k}}}$ the dipole matrix elements, $\ket{u_m^{\vec{k}}}$ the cell-periodic part of the Bloch function $\ket{\phi_m^{\vec{k}}}$,  $\rho_{mn}^{\vec{k}}$ the density matrix elements, $T_2$ the dephasing time, and $\vec{\berryc}_n^{\vec{k}}\equiv \vec{d}_{nn}^{\vec{k}}$ the Berry connections.

The total current can be split into the interband and intraband contributions by
\begin{subequations}
  \label{eq:theory_sbe_2}
  \begin{align}
    \vec{j}_{\text{ter}}(t)
    = & - \sum_{\vec{K}} \sum_{m\ne n} \rho_{nm}^{\vec{K}}(t) \vec{p}_{mn}^{\vec{K}+\vec{A}(t)} \label{eq:theory_sbe_2a} \\
    \vec{j}_{\text{tra}}(t)
    = & - \sum_{\vec{K}} \sum_{n} \rho_{nn}^{\vec{K}}(t) \vec{p}_{nn}^{\vec{K}+\vec{A}(t)}, \label{eq:theory_sbe_2b}
  \end{align}
\end{subequations}
with $\vec{p}_{mn}^{\vec{k}} = \bra{\phi_m^{\vec{k}}} \vop{p} \ket{\phi_n^{\vec{k}}}$ the momentum matrix elements, and the summation over $\vec{K}$ is over the whole BZ (throughout the text). 
The HHG spectrum is taken as the modulus squares of the Fourier transforms of the currents, after weighting by a window function.

Throughout this work, we make use of the two-band approximation with an initially filled valence band labelled ``v'' and an empty conduction band labelled ``c''.
For notational convenience, we henceforth use $\vec{d}^{\vec{k}}\equiv \vec{d}^{\vec{k}}_{cv}$ for the transition dipole and $\omega_g^{\vec{k}}\equiv E_c^{\vec{k}}-E_v^{\vec{k}}$ for the band gap.

\subsection{Extended semiclassical picture} \label{sec:theory_semiclass}
In this subsection, we go over the semiclassical models used in this work. We start by obtaining the saddle point equations from the SBEs, and then describe our extended recollision model. The case for linearly-polarized fields was partly discussed in \cite{Yue2020PRL}. More info on the details of the derivation can be found in Appendix~\ref{sec:app_derivs}.

\subsubsection{Saddle-point equations}
For the laser pulses and systems considered in this work, the conduction band population during the laser is small [$\rho_{vv}^{\vec{k}}(t)-\rho_{cc}^{\vec{k}}(t)\approx 1$], such that solutions to Eq.~\eqref{eq:theory_sbe_1} can formally be written down. The interband spectrum is $\vec{j}_{\text{ter}}(\omega) = \int_{-\infty}^{\infty} dt e^{i\omega t} \vec{j}_{\text{ter}}(t)$, where the Cartesian ($\pol=\{x, y, z\}$) current components in the fixed frame are (see the derivation in Appendix~\ref{sec:app_derivs_1})
\begin{equation}
  \label{eq:theory_sfa_1}
  \begin{aligned}
    j_{\text{ter},\pol}(t)
    & = \sum_{\vec{k}} R^{\vec{k}}_\pol  \int^t
     T^{\vec{\kappa}(t,s)}
    e^{-iS^\pol(\vec{k}, t, s)} ds
    + \text{c.c.}
  \end{aligned}
\end{equation}
with $T^{\vec{\kappa}(t,s)}=\left|\vec{F}(s)\cdot\vec{d}^{\vec{\kappa}(t,s)}\right|$ the transition matrix element, $R^{\vec{k}}_\pol =  \omega_{g}^{\vec{k}}| {d}_{\pol}^{\vec{k}} |$ the recombination dipole, and $\vec{\kappa}(t, t') = \vec{k} - \vec{A}(t) + \vec{A}(t')$. The times $s$ and $t$ can be interpreted as the excitation and emission times, respectively.
The accumulated phase in Eq.~\eqref{eq:theory_sfa_1} is (dephasing ignored)
\begin{equation}
  \label{eq:theory_sfa_2}
  \begin{aligned}
    S^\pol(\vec{k}, t, s)
    =& \int_s^t \left[ \omega_g^{\vec{\kappa}(t, t')} + \vec{F}(t')\cdot\Delta\berryc^{\vec{\kappa}(t, t')} \right] dt' \\
    & + \alpha^{\vec{k},\pol}
    - \beta^{\vec{\kappa}(t,s)}
  \end{aligned}
\end{equation}
with $\Delta \berryc^{\vec{k}} \equiv \berryc_c^{\vec{k}} - \berryc_v^{\vec{k}}$, $\alpha^{\vec{k},\pol} \equiv\arg({d}_\pol^{\vec{k}})$ the transition-dipole phases (TDPs), and $\beta^{\vec{\kappa}(t,s)}\equiv \arg[\vec{F}(s)\cdot \vec{d}^{\vec{\kappa}(t,s)}]$. The saddle point conditions for $S^\pol(\vec{k},t,s)-\omega t$ read
\begin{subequations}
  \label{eq:theory_sfa_3}
  \begin{align}
    \omega_g^{\vec{\kappa}(t, s)} + \vec{F}(s) \cdot \ccom^{\vec{\kappa}(t, s)}
    & = 0, \label{eq:theory_sfa_3a}
    \\
    \Delta \vec{R}^\pol \equiv \Delta\vec{r} - \com^{\vec{k},\pol} + \ccom^{\vec{\kappa}(t, s)}
    & = \vec{0}, \label{eq:theory_sfa_3b}
    \\
    \omega_g^{\vec{k}}
      + \vec{F}(t) \cdot \left[ \ccom^{\vec{\kappa}(t, s)} + \Delta\vec{r}  \right]
    & = \omega, \label{eq:theory_sfa_3c}
  \end{align}
\end{subequations}
where the electron-hole separation vector and group velocities are
\begin{subequations}
  \label{eq:theory_sfa_4}
  \begin{align}
    \Delta \vec{r} \equiv
    & \int_s^t \left[ \vec{v}_c^{\vec{\kappa}(t, t')} - \vec{v}_v^{\vec{\kappa}(t, t')} \right] dt'   \label{eq:theory_sfa_4a} \\
    \vec{v}_n^{\vec{\kappa}(t, t')} \equiv
    & \nabla_{\vec{k}}E_n^{\vec{\kappa}(t, t')} + \vec{F}(t') \times \vec{\Omega}_n^{\vec{\kappa}(t, t')},   \label{eq:theory_sfa_4b}
  \end{align}
\end{subequations}
with the Berry curvature $\vec{\Omega}_n^{\vec{k}} \equiv \nabla_{\vec{k}} \times \berryc_n^{\vec{k}}$, and 
\begin{subequations}
  \label{eq:theory_sfa_5}
  \begin{align}
    \com^{\vec{k},\pol} \equiv \Delta \berryc^{\vec{k}} - \nabla_{\vec{k}}\alpha^{\vec{k},\pol} \label{eq:theory_sfa_5a} \\
    \ccom^{\vec{k}} \equiv \Delta \berryc^{\vec{k}} - \nabla_{\vec{k}}\beta^{\vec{k}} \label{eq:theory_sfa_5b}.
  \end{align}
\end{subequations}
Note that in our notation, $\mu$ used as a subscript points to a scalar quantity, while $\mu$ used as a superscript correspond to a function variable: for example, $\com^{\vec{k},\mu}$ is a vector that depends on $\mu$.

Equations~\eqref{eq:theory_sfa_3a}-\eqref{eq:theory_sfa_3c} can be interpreted by the following three steps in the interband HHG process: at time $s$, an electron-hole pair is created by tunnel excitation at the crystal momentum $\vec{k}_0\equiv\vec{\kappa}(t, s)$; the laser accelerates the electron and hole with the instantaneous group velocities $v_n^{\vec{\kappa}(t,t')}$; the electron and hole recollide at time $t$ with final crystal momentum $\vec{k}$ and relative distance $\Delta\vec{r}$, with the simultaneous emission of high-harmonics with energy $\omega$.

The saddle-point equations first proposed by Vampa and co-workers \cite{Vampa2014, Vampa2015PRB, Vampa2015Nat} include only the first term on the left-hand sides of Eqs.~\eqref{eq:theory_sfa_3a}-\eqref{eq:theory_sfa_3c}. The above equations includes additionally (i) laser-dressing of the bands with $\vec{F}\cdot\ccom^{\kappa(t, s)}$ in Eqs.~\eqref{eq:theory_sfa_3a} and \eqref{eq:theory_sfa_3c}; (ii) the anomalous velocity term \cite{Xiao2010} in Eq.~\eqref{eq:theory_sfa_4a} involving the Berry curvatures; (iii) a shift of the recollision condition in Eq.~\eqref{eq:theory_sfa_3b} and (iv) the possibility of an additional nonzero electron-hole-pair polarization energy at recollision (\ehPE) $\vec{F}\cdot \Delta\vec{r}$ in Eq.~\eqref{eq:theory_sfa_3c}. Physically, the \ehPE{} constitutes the potential energy of the electric dipole comprised of the positively-charged hole and negatively-charged electron at the time of recollision. We note that in systems with inversion and time-reversal symmetries, the Berry curvatures are zero.
We mention that equivalent equations to Eq.~(5) appear in \cite{Li2019phase}, but with the strict constraint $\Delta\vec{r}=\vec{0}$ such that the \ehPE{} is zero.

Note that for the all physics to be relevant and consistent, under an arbitrary ``structure''-gauge transformation $\ket{u_n^{\vec{k}}} \rightarrow \ket{u_n^{\vec{k}}}e^{i\varphi_n^\vec{k}}$ ($\varphi_n^{\vec{k}}\in\mathbb{R}$), Eq.~\eqref{eq:theory_sfa_3} should remain unchanged.
While the individual terms in the right-hand sides of Eq.~\eqref{eq:theory_sfa_5} generally depend on the gauge-choice, the composed quantities, $\com^{\vec{k}}_\pol$ and $\ccom^{\vec{k}}_\pol$ are shown to be gauge invariant in Appendix~\ref{sec:app_derivs_2}. The gauge invariance of the saddle-point equations in Eq.~\eqref{eq:theory_sfa_3} then trivially follows.

In many studies of solid-state HHG, linearly polarized drivers $\vec{F}(t)=F(t)\vec{\hat{e}}_\|$ are used, in which case Eq.~\eqref{eq:theory_sfa_3} reduces to
\begin{subequations}
  \label{eq:theory_sfa_6}
  \begin{align}
    \omega_g^{\vec{\kappa}(t,s)} + \vec{F}(s) \cdot \com^{\vec{\kappa}(t,s)}_\parallel
    & = 0, 
    \\
    \Delta \vec{R}^\pol \equiv \Delta\vec{r} - \com^{\vec{k},\pol} + \com^{\vec{\kappa}(t,s),\|}
    & = \vec{0}, 
    \\
    \omega_g^{\vec{k}}
      + \vec{F}(t) \cdot \left[ \com^{\vec{\kappa}(t,s), \|} + \Delta\vec{r}  \right]
    & = \omega, 
  \end{align}
\end{subequations}
with $\pol=\|,\perp_1,\perp_2$ where $\vec{\hat{e}}_{\perp_1}$ and $\vec{\hat{e}}_{\perp_2}$ are unit vectors perpendicular to $\vec{\hat{e}}_\|$, and we used $\beta^{\vec{\kappa}(t,s)}=\alpha^{\vec{\kappa}(t,s),\|}+\arg[F(s)]$ in the derivation.

\subsubsection{Extended recollision model}
We solve the saddle-point equations \eqref{eq:theory_sfa_3} semiclassically, in an extension to the ``original'' recollision model, and will refer to it as the extended recollision model (ERM). We note that we first introduced this method in Ref.~\cite{Yue2020PRL}, for a linearly polarized field.
Since the bandgaps in semiconductors and insulators are never zero, we neglect to solve Eq.~\eqref{eq:theory_sfa_3a}. Instead, we choose to consider the electron-hole creation at an initial crystal momentum, $\vec{k}_0 \equiv \vec{\kappa}(t,s)$, taken to be close to or at a high-symmetry point. We then integrate the group velocities in Eq.~\eqref{eq:theory_sfa_4b} to obtain the classical motions of the electron and hole, with the time-dependent crystal momentum given by
\begin{equation}
  \label{eq:theory_sfa_6}
  \vec{\kappa}(t,t') = \vec{k}_0 + \vec{A}(t') - \vec{A}(s), \quad t' \in [s, t].
\end{equation}

During the propagation, we calculate the ``generalized'' electron-hole distance vector $\Delta \vec{R}^\pol$ in Eq.~\eqref{eq:theory_sfa_3b}, and a returning trajectory is said to have recollided at time $t'=t$ if: (i) $\norm{\Delta \vec{R}^\pol}$ as a function of $t'$ is at a local minimum and (ii) the $\norm{\Delta \vec{R}^\pol}<R_0$ requirement is fulfilled, where $R_0$ is a preset recollision threshold value.
In our calculations, we will use $R_0\in[15,100]$. The $R_0$ is chosen such that the semiclassical results agree with the time-frequency profiles. Using a larger $R_0$ value for a given calculation will make the features in the recollision-energy-vs-time spectrum broader, but the same qualitative trend remains. As also discussed in \cite{Yue2020PRL}, the minimum $R_0$ that yields reasonable agreement with the SBE calculations is a measure of the effective size of the recolliding quantum wave packet.
We set $R_0=30$ unless indicated otherwise.  If a trajectory has recollided, we record $s$, $t$ and the recollision energy $\omega(\vec{k}_0,s,t)$ in Eq.~\eqref{eq:theory_sfa_3c}.
We initiate trajectories for times spanning an optical cycle (o.c.) $s\in[-T,0]$, and propagate each trajectory maximally up two optical cycles after tunneling $t\in[s, s+2T]$. For each trajectory, we count up to a maximum of 3 recollisions. However, unless indicated otherwise, we present results for the first recollision. Often, we perform calculations for all $\vec{k}_0$s in a disc with radius $\Delta k$ around a high-symmetry point.
%
We note that allowing electron tunneling at different $\vec{k}_0$ points away from the minimum band gap is inherently distinct from the summation over all $\vec{k}$ points in the expressions for the total current in Eqs.~\eqref{eq:theory_sbe_2} and \eqref{eq:theory_sfa_1}. In the former case, we attempt to find all the stationary-phase points \{$\vec{k}$, $s$, $t$\} that most contribute to the integral~\eqref{eq:theory_sfa_1}.

We formally define an imperfect recollision as having a nonzero electron-hole distance $\norm{\Delta\vec{R}^\pol} \ne 0$, i.e. whenever the electron and hole centers do not exactly spatially reencounter each other. Note that the electron-hole pair will get driven further apart spatially whenever the direction of the time-dependent crystal momentum [and thus $\vec{A}(t')$] in Eq.~\eqref{eq:theory_sfa_6} is not along the group velocities in Eq.~\eqref{eq:theory_sfa_4b}, and in such cases the imperfect recollisions could be important. One can thus control and force such recollisions by either tuning the laser or studying different materials, e.g. by using elliptical drivers or considering materials with large Berry curvatures. Also, since the group velocities are the gradient of the band dispersions, tunneling at a reciprocal point that is not the minimum band gap can also lead to imperfect recollisions \cite{Yue2020PRL}, even when using linearly polarized driving fields.

It should be mentioned that recent progress \cite{Navarrete2019, Uzan2020, Parks2020} has been made towards solving the saddle point equations \eqref{eq:theory_sfa_3} and performing the stationary phase approximation on the integral in Eq.~\eqref{eq:theory_sfa_1}. However, these studies present a monumental task, even in reduced dimensionalities and without the extra terms involving $\ccom^{\vec{\kappa}(t,s)}$ and $\com^{\vec{k},\pol}$. It also remains to be seen whether such formalisms can treat electron-hole-pair creation at different symmetry points in the BZ.

\subsection{Electron and hole wave-packet analysis} \label{sec:theory_wp}
We present here a quantum wave packet method that is able to provide additional details on the spatially extended nature of the imperfect recollisions, by explicitly constructing the real-space electron and hole wave packets for a specific semiclassical trajectory. We label this method, which was first applied in our previous work \cite{Yue2020PRL} for the electron wave packet, as the wave-packet trajectory (WPT) method.

For concreteness, consider a semiclassical electron trajectory in the conduction band that tunneled at time $s$ and reciprocal coordinate $\vec{k}_0=\vec{K}_0-q\vec{A}(s)$, where $q=-1$ is the electron charge and $\vec{k}_0$ ($\vec{K}_0$) is the crystal momentum in the fixed (moving) frame. We expand the real-space electron wave packet in the Houston-state basis
\begin{align}
  \label{eq:theory_wp_1}
    \Psi_e(\vec{r}, t) = \sum_{\vec{K}}a_e^{\vec{K}}(t)h_c^{\vec{K}}(\vec{r},t), 
\end{align}
where $\abs{a_c^{\vec{K}}(s)}^2$ is chosen to be a Gaussian centered at $\vec{K}_0$, with a full width at half maximum (FWHM) estimated by Zener-tunneling as described in Appendix~\ref{sec:app_derivs_3} and Eq.~\eqref{eq:app_tun_1}. The Houston states \cite{Houston1940, Krieger1986} are related to the accelerated Bloch states
\begin{equation}
  \label{eq:theory_wp_2}
  h_c^{\vec{K}}(\vec{r}, t) = e^{iq\vec{A}(t)\cdot\vec{r}} \phi_c^{\vec{K} - q\vec{A}(t)}(\vec{r}).
\end{equation}
Inserting Eq.~\eqref{eq:theory_wp_1} into the time-dependent Schr\"odinger equation, and neglecting coupling to the other bands, leads to the equations of motion:
\begin{equation}
  \label{eq:theory_wp_3}
  i \dot{a}_e^{\vec{K}} (t)
  = \left[ E_c^{\vec{K}-q\vec{A}(t)}
    - q \vec{F}(t)\cdot \vec{\berryc}_{c}^{\vec{K}-q\vec{A}(t)} \right] a_e^{\vec{K}}(t).
\end{equation}
We thus propagate Eq.~\eqref{eq:theory_wp_3} starting from time $s$, and at desired time intervals calculate the real-space wave packet using Eq.~\eqref{eq:theory_wp_1}.
More details on the evaluation of the wave packet is given in Appendix~\ref{sec:app_derivs_4}.
With access to the real-space wave packet, the observables such as the expectation values $\left<\vec{r}\right>(t)$ and standard deviations $\sigma_{\pol}=\sqrt{\left<\op{\pol}^2\right>-\left<\op{\pol}\right>^2}$, with $\op{\pol}\in\left\{\op{x},\op{y},\op{z}\right\}$, can be calculated. For better visualization of the width, we define the FWHM-like $\bar{\sigma} \equiv \sqrt{2\log 2}(\sigma_x+\sigma_y)$.
We note that since the initial phase of $a_c^{\vec{K}}(s)$ is unknown (we set the phase to zero), the width of the wave packet will have a dependence on the phase of the structure gauge chosen for $\ket{\phi_c^{\vec{K}}}$. However, since the twisted parallel transport gauge has optimally smooth Bloch states \cite{Vanderbilt2018}, we expect this dependence to be small.

A hole is left behind in the valence band when an electron tunnels from the valence to the conduction band. Seen as a quasiparticle, the hole has positive charge $q_h=1$, and satisfies $\vec{K}_h=-\vec{K}$ (total crystal momentum conservation) and $E_h^{\vec{k}}=-E_v^{\vec{k}}$. The corresponding equations for the hole wave packet $\Psi_h(\vec{r},t)$ is then obtained from Eqs.~\eqref{eq:theory_wp_1} and \eqref{eq:theory_wp_3} by substituting in the equations $a_e\rightarrow a_h$, $h_c\rightarrow h_v$, $\vec{K}\rightarrow \vec{K}_h$, $q\rightarrow q_h$ and $E_c\rightarrow E_h$.

Finally, it should be noted that the construction of the quantum wave packet in Eqs.~\eqref{eq:theory_wp_1} and \eqref{eq:theory_wp_2} requires the knowledge of the Bloch wave functions $\phi_n^{\vec{k}}(\vec{r})$ which sometimes can be hard to obtain.

\section{Ellipticity dependency in a bulk solid} \label{sec:zno2d}

Recently, the ellipticity-dependence of HHG in bulk solids has attracted both theoretical and experimental attention \cite{Ghimire2011, You2017, Ndabashimiye2016, Liu2017, Tancogne-Dejean2017, Yoshikawa2017, Zurron-Cifuentes2018, Zhang2019, Hollinger2021nanomat}. In contrast to HHG in gases, where the HHG yield falls off with increasing ellipticity, HHG in solids exhibits nontrivial ellipticity dependence where the harmonic yield can increase with increasing ellipticity. In this section, we investigate the ellipticity-dependence of HHG in a generic bulk-solid system with the minimum band gap at the $\Gamma$ point. We consider a model for bulk ZnO, using a two-band approximation and neglecting the Berry connections, Berry curvatures, and TDPs. As we will show below, this treatment allows a detailed and quantitative understanding of the recolliding trajectories and emission dynamics in a generic bulk solid.

\subsection{Generic bulk solid: ZnO model} \label{sec:zno2d_model}
\begin{figure}
  \centering
  \includegraphics[width=0.48\textwidth, clip, trim=0 0cm 0 0cm]{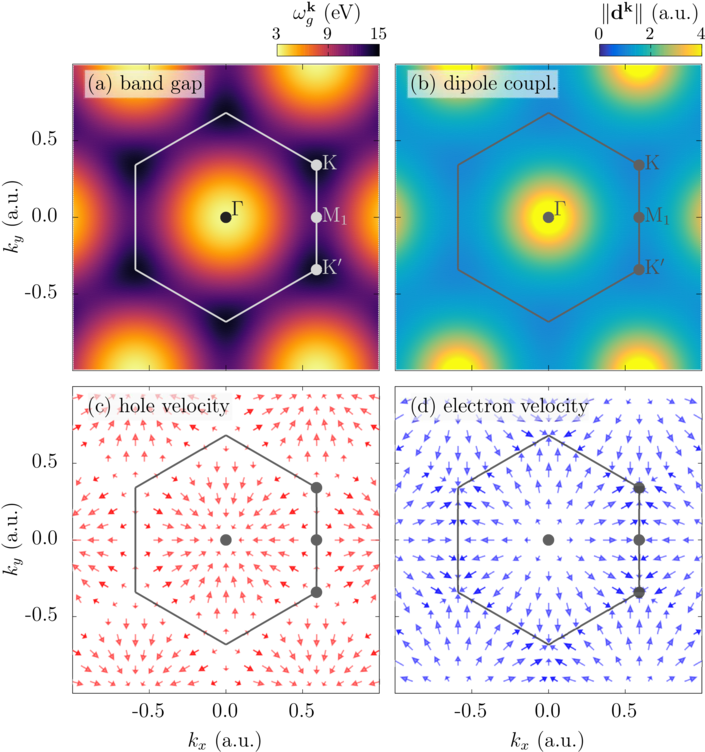}
  \caption{(a) Band gap of ZnO in the plane containing the high symmetry points $\Gamma$, $K$ and $M$. (b) Norm of the transition dipole moment. (c) Group velocities of holes in the valence band. (d) Group velocities of the electrons in the conduction band. The hexagon in the plots guides the eye and traces the first BZ.}
  \label{fig:zno2d_struc}
\end{figure}

For the band structure of wurtzite ZnO, we consider the plane containing the $\Gamma$, $K$ and $M$ high-symmetry points. The band structure is obtained using the analytical model
\begin{subequations}
  \label{eq:zno2d_struc_1}
  \begin{align}
    & E_n^{\vec{k}}= u^{-1} \left[ t_n \sqrt{f^{\vec{k}}+q_n} + t' f^{\vec{k}} + p_n \right],  \label{eq:bulk_struc_1a} \\
    & f^{\vec{k}} = 2\cos(ak_y) + 4\cos\left[ \frac{1}{2} ak_y \right] \cos\left[\frac{\sqrt{3}}{2} ak_x\right], \label{eq:zno2d_struc_1b}
  \end{align}
\end{subequations}
with the fitted parameters $t_v=2.38$, $t_c=-2.38$, $q_v=4.0$, $q_c=3.3$, $t'=-0.020$, $p_v=-7.406$, $p_c=10.670$, $u=27.1$. Our model is adapted from Ref.~\cite{Zhang2019}, but now using the real lattice constant of $a=6.14$ for ZnO. The $\vec{k}$-dependent band gap shown in Fig.~\ref{fig:zno2d_struc}(a) is seen to exhibit hexagonal symmetry with the minimum band gap at $\Gamma$ $\omega_g^{\Gamma}=3.3$ eV. The transition dipole is taken to be real and approximated by \cite{Vampa2014, McDonald2015}
\begin{equation}
  \label{eq:zno2d_struc_2}
  d_x^{\vec{k}}=d_y^{\vec{k}}=\sqrt{\frac{K}{2(\omega_g^{\vec{k}})^2}}
\end{equation}
with the Kane parameter $K=0.302$. Fig.~\ref{fig:zno2d_struc}(b) shows the dipole magnitude, with the obvious maximum at $\Gamma$.


The group velocities for the valence and conduction bands are plotted as vector fields in Figs.~\ref{fig:zno2d_struc}(c) and Figs.~\ref{fig:zno2d_struc}(d), respectively. For the hole (electron) group velocities, the $\Gamma$ ($K$) point acts as a sink with the vectors pointing towards it, while the $K$ ($\Gamma$) point acts as a source.

Note that even though a hexagonal BZ is visualized in Fig.~\ref{fig:zno2d_struc}, in the actual calculations we use a Monkhorst-Pack mesh spanned by the reciprocal vectors $\vec{b}_1=2\pi( 3^{-\frac{1}{2}}\uv{e}_x + \uv{e}_y )/a$ and $\vec{b}_2=2\pi( 3^{-\frac{1}{2}}\uv{e}_x - \uv{e}_y )/a$.

\subsection{Driver ellipticity dependence of HHG in ZnO} \label{sec:zno2d_hhg}
\begin{figure}
  \centering
  \includegraphics[width=0.48\textwidth, clip, trim=0 0cm 0 0cm]{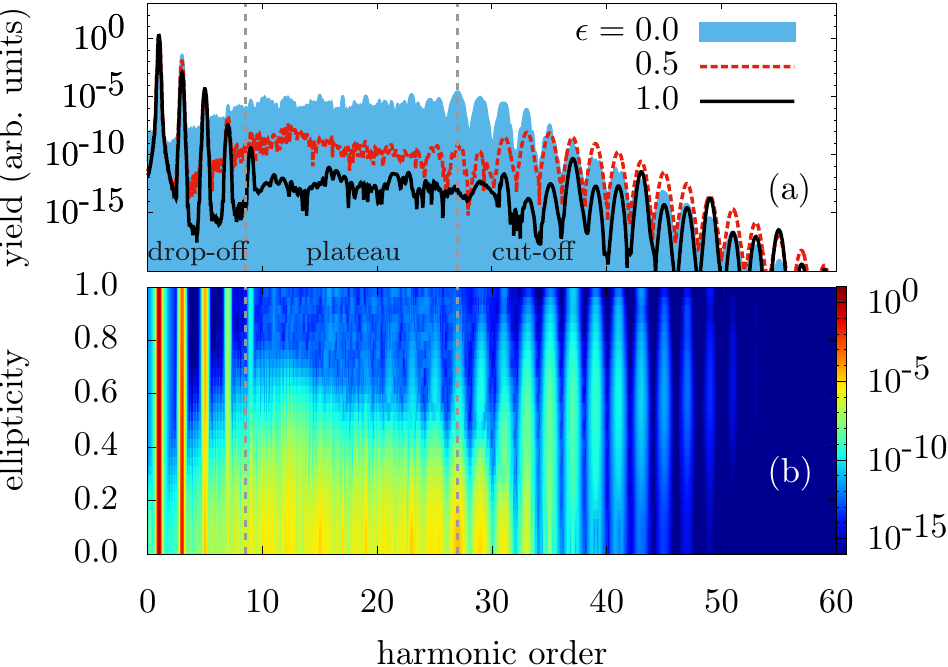}
  \caption{High-harmonic spectra of ZnO driven by elliptically polarized light (major axis of ellipse along $\Gamma$-M). The laser parameters are $\omega_0=0.0142$ ($\lambda=3200$ nm), $A_0=0.35$ and $\tau=106.7$ fs, with dephasing $T_2=10$ fs. The first vertical dashed line at $\sim$\hh 8.5 traces the minimum band-gap energy $\omega_g^{\Gamma}$ and separates the drop-off region from the plateau region; the second vertical line at \hh 27 guides and eye and approximately separates the plateau and cut-off regions.}
  \label{fig:zno2d_ellip}
\end{figure}

We irradiate the bulk with elliptically-polarized vector potentials of the form
\begin{equation}
  \label{eq:zno2d_sbe_1}
  \vec{A}(t)= \frac{A_0g(t)}{\sqrt{1+\epsilon^2}} \left[
    \sin(\omega_0 t)\uv{e}_x +  \epsilon \cos(\omega_0 t) \uv{e}_y
  \right],
\end{equation}
where $\ellip$ is the ellipticity, $\omega_0$ is the carrier frequency, $F_0 =\omega_0A_0$ is the electric field maximum, and the pulse envelope is on the form $g(t)=\cos^2[\pi t/(2\tau)]$ with $t\in[-\tau,\tau]$.
For our calculations in this section, we choose $\omega_0=0.0142$ ($\lambda=3200$ nm), $A_0=0.35$ and $\tau=106.7$ fs. We note that Eq.~\eqref{eq:zno2d_sbe_1} describes an elliptically polarized field with major axis along the $\Gamma-M_1$ ($\uv{e}_x$) direction. Simulations with the ellipse major axis along $\Gamma-K$ yields nearly indistinguishable results from those in Fig.~\ref{fig:zno2d_ellip} and will not be discussed further in this work.


Figure~\ref{fig:zno2d_ellip}(a) shows the HHG spectrum for three different ellipticities: $\ellip=0$ (linear polarization), $\ellip=0.5$, and $\ellip=1.0$ (circular polarization).
The HHG spectra are seen to be divided into three regions by the vertical lines at $\omega_g^{\Gamma}$ and harmonic 27 (\hh 27): a \textit{drop-off} region, a \textit{plateau} region and a \textit{cut-off} region. For $\ellip=0.5$, the harmonic intensity in the plateau region is reduced by up to 5 orders of magnitudes compared to $\ellip=0$, while in the cut-off region the harmonic yield is actually increased going from $\ellip=0$ to $\ellip=0.5$. The shape of the spectrum for circular polarization is qualitatively similar to the $\epsilon=0.5$ case, but with an overall decrease in yield in the plateau and cut-off regions.

Figure~\ref{fig:zno2d_ellip}(b) shows a more complete analysis with the HHG spectrum scanned over the ellipticities $\ellip\in[0,1]$. We focus our attention on the harmonics with energies above the minimum band gap $\omega_g^{\Gamma}$ where the interband harmonics dominate. In the plateau region, a monotonic decrease of yield with increasing $\ellip$ is evident, with the yield almost vanishing at $\ellip=1$. Such a behavior is similar to the ellipticity dependence of HHG in gases. The cut-off region in Fig.~\ref{fig:zno2d_ellip}(b), however, exhibits anomalous ellipticity-dependence, with relatively large yields between $\ellip=0$ and $\ellip=1$. Qualitatively similar ellipticity-dependencies were reported in Refs.~\cite{Li2019, Zhang2019}.

\subsection{Emission profiles and semiclassical analysis for $\Gamma$} \label{sec:zno2d_gamma}


\begin{figure}
  \centering
  \includegraphics[width=0.48\textwidth, clip, trim=0 0cm 0 0cm]{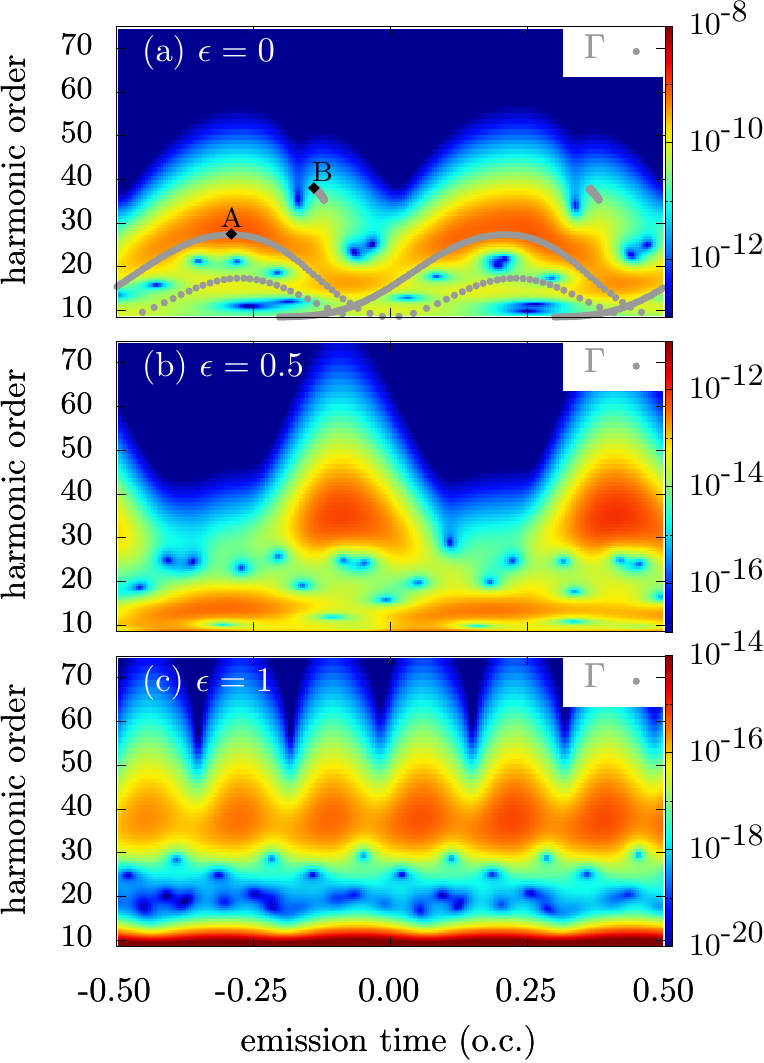}
  \caption{Time-frequency emission profiles (colormaps) of the harmonics for three different ellipticities: (a) $\ellip=0$; (b) $\ellip=0.5$ and (c) $\ellip=1.0$. The black lines are semiclassical ERM results with tunneling initiated at $\vec{k}_0=\Gamma$, counting the first two recollisions, and assuming the recollision thresholds $R_0=30, 60, 100$ for $\ellip=0, 0.5, 1$, respectively. Note that semiclassical recollisions with tunneling from the $\Gamma$ point are only observed for the linearly polarized driver in (a). A and B mark different classes of trajectories as discussed in the text.}
  \label{fig:zno2d_cwt}
\end{figure}

The character and periodicity of the harmonic time-frequency emission profiles also depend strongly on the ellipticity, as illustrated in Fig.~\ref{fig:zno2d_cwt}.
For $\ellip=0$ in Fig.~\ref{fig:zno2d_cwt}(a), the profile exhibits half-cycle periodicity, with the most prominent feature exhibiting a peak at around \hh 27 and emitted at $-0.29$ o.c. Overall, it resembles a typical time-frequency profile from HHG in gases, where every energy below the maximum is emitted twice, corresponding to the \textit{short} and \textit{long} trajectories. In contrast, the time-frequency profile for $\ellip=0.5$ in Fig.~\ref{fig:zno2d_cwt}(b) looks entirely different: the highest-order harmonics in the cut-off region are much more dominant and energetic, exhibiting a characteristic broad \textit{triangular} structure, and with the emission time shifted to $\sim-0.1$ o.c. The harmonics emitted with energies corresponding to the plateau region (\hh 9 to \hh 21) are shifted in time by a quarter cycle compared to the triangular structure. The time-frequency profile for $\ellip=1$ in Fig.~\ref{fig:zno2d_cwt} shows six burst of light during each o.c. -- a clear reflection of the six-fold rotational symmetry of the BZ (see Fig.~\ref{fig:zno2d_struc}).


\begin{figure}
  \centering
  \includegraphics[width=0.48\textwidth, clip, trim=0 0cm 0 0cm]{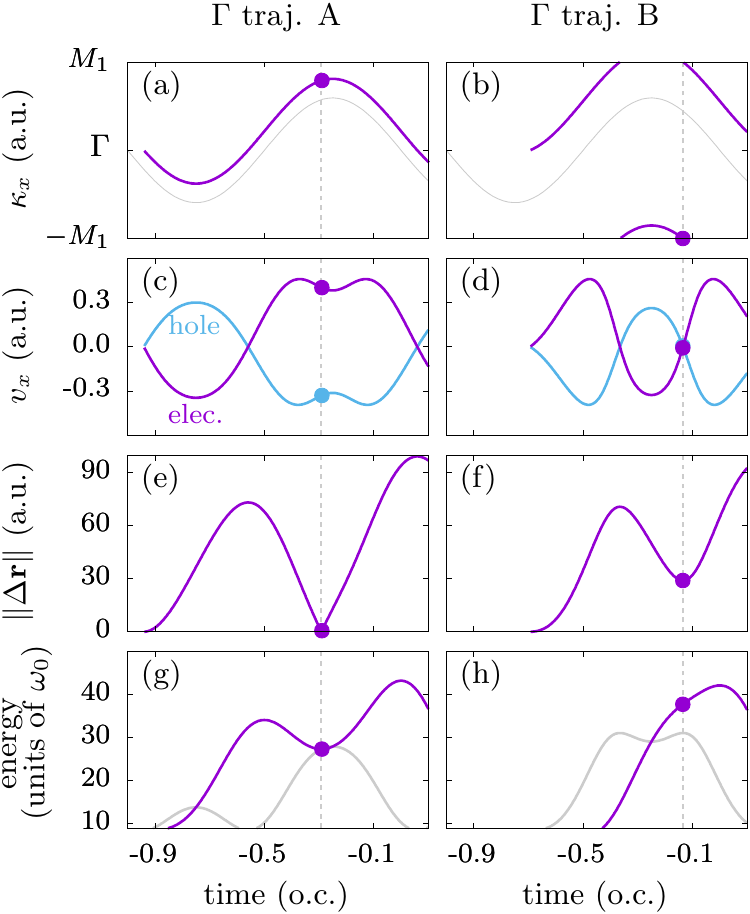}
  \caption{Two different semiclassical trajectories for the linearly polarized driver with $\Gamma$ as the tunnel point. The recollision event is marked with the filled circle in each panel. Left panels: trajectory that (perfectly) recollide at $t=-0.29$ o.c. with recollision energy $\omega=27\omega_0$; right panels: trajectory that (imperfectly) recollide at $t=-0.14$ o.c. with recollision energy $\omega=38\omega_0$. The gray lines in (a) and (b) show $A_x(t)$. In (g) and (h), the gray lines show the energies without the \ehPE{}, while the purple lines show the total energies.}
  \label{fig:zno2d_traj1}
\end{figure}


We first analyze the emission profiles for the linearly-polarized case, by using the ERM in Sec.~\ref{sec:theory_semiclass} and assuming that tunneling occurs at the minimum band gap $\Gamma$. The emission times for individual trajectories are shown in Fig.~\ref{fig:zno2d_cwt}(a) by the gray dots. The agreement with the colormap is quite good, with the semiclassical results reproducing the emission profiles during each half-cycle. Even the peculiar structure at $\sim$\hh 38 is captured by the semiclassical model. The very different emission profiles of the trajectories labeled A and B in Fig.~\ref{fig:zno2d_cwt}(a) suggest that they belong to different classes of trajectories. This is further illustrated in Fig.~\ref{fig:zno2d_traj1}, where we in the left panels consider trajectory A. Figure~\ref{fig:zno2d_traj1}(a) shows the reciprocal-space motion: the electron-hole pair is created at the $\Gamma$ point at time $s=-0.941$ o.c., and afterward is driven by the vector potential according to Eq.~\eqref{eq:theory_sfa_6}. The time-dependent crystal momentum initially moves toward $-M_1$, and later changes direction when the vector potential changes direction; it never moves beyond the BZ boundaries (at $\pm M_1$). In Fig.~\ref{fig:zno2d_traj1}(c), the electron group velocity is negative (positive) when $k_x<0$ ($k_x>0$), which is also observed in Fig.~\ref{fig:zno2d_struc}(d). The electron and hole undergo a perfect recollision ($\norm{\Delta \vec{r}}=0$) at time $-0.29$ o.c.  in Fig.~\ref{fig:zno2d_traj1}(e), and consequently the recollision energy in Fig.~\ref{fig:zno2d_traj1}(g) with and without the \ehPE{} is the same. Trajectories of class A in Fig.~\ref{fig:zno2d_cwt}(a) are thus similar to the ones in HHG in gases, consisting of short and long trajectories.

Consider now the special trajectory labelled B in Fig.~\ref{fig:zno2d_cwt}(a). After tunneling at $\Gamma$, the crystal momentum goes beyond the BZ-boundary [Fig.~\ref{fig:zno2d_traj1}(b)], where electron and hole group velocities abruptly change sign [Fig.~\ref{fig:zno2d_traj1}(d)], and undergo a Bragg reflection. Consequently, the electron and hole only imperfectly recollide in real space, with $\norm{\Delta\vec{r}}\approx 30$ shown in Fig.~\ref{fig:zno2d_traj1}(f). The resulting extra \ehPE{} contributes $\sim 7 \omega_0$ which is added to the total energy of the emitted harmonics in Fig.~\eqref{fig:zno2d_traj1}(h). Bragg reflections can thus lead to imperfect recollisions in bulk solids even when using linearly-polarized drivers. Note that the effects of the Bragg reflection on HHG in solids have been investigated in several previous works \cite{Ghimire2011, Ghimire2012PRA, Hawkins2015, Du2018PRA, Zhang2019}.

In contrast to the linear-polarization case, the elliptically polarized fields do not lead to any recollisions initiated from the $\Gamma$ point, as shown in Figs.~\ref{fig:zno2d_cwt}(b) and ~\ref{fig:zno2d_cwt}(c). To ensure that this is not just due to a larger recollision distance, the recollision thresholds has been relaxed from $R_0=30$ at $\ellip=0$ to $R_0=60$ at $\ellip=0.5$ and $R_0=100$ at $\ellip=1$. Including only trajectories initiated at the minimum band is thus insufficient for the description of HHG with elliptically polarized drivers in bulk solids.

\subsection{Full semiclassical picture - effect of the full BZ} \label{sec:zno2d_full}


\begin{figure}
  \centering
  \includegraphics[width=0.5\textwidth, clip, trim=0 0cm 0 0cm]{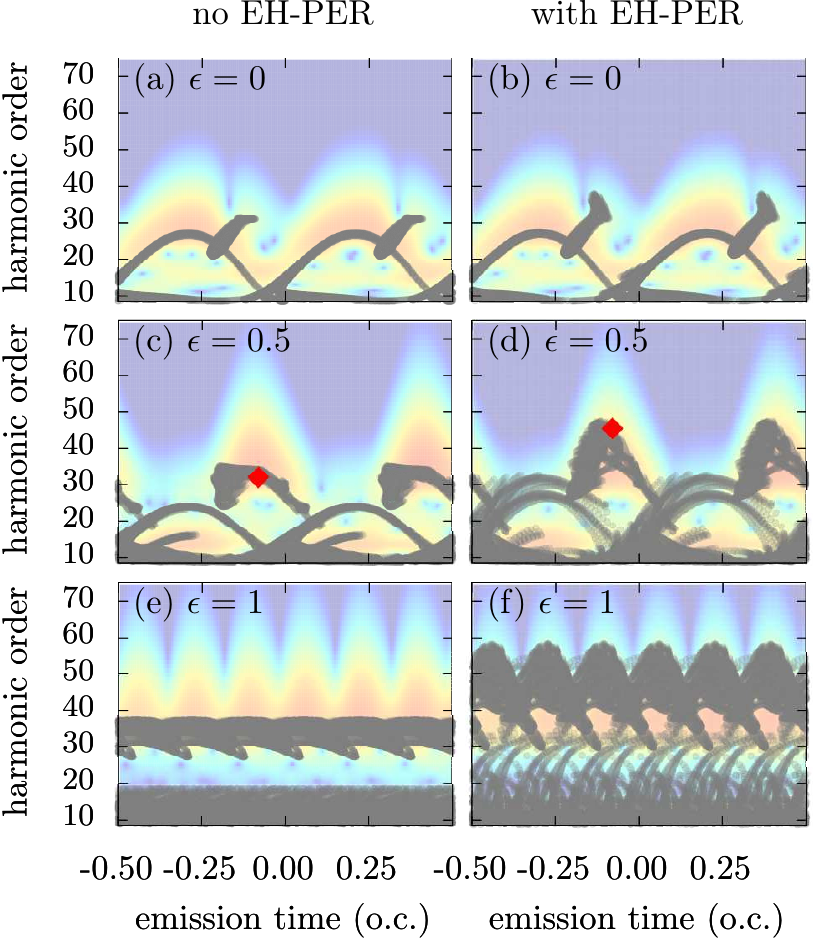}
  \caption{Semiclassical recollision energies versus recollision times obtained using the ERM (gray dots) with tunneling initiated for all $\vec{k}_0$ in a disc around $\Gamma$. Left (right) panels show the semiclassical results without (with) the inclusion of the \ehPE{} $\vec{F}(t)\cdot\Delta\vec{r}$ in Eq.~\eqref{eq:theory_sfa_3c}, while different row panels correspond to different ellipticities $\ellip$. For $\ellip=0, 0.5, 1$, we chose $\Delta k=0.15, 0.2, 0.25$ and $R_0=30, 60, 100$, respectively (see text). The background show the time-frequency profiles from Fig.~\eqref{fig:zno2d_cwt} using the same color scale. The red diamonds in (c) and (d) highlight the trajectories that tunnel at $s=-0.83$ o.c. and recollide at $t=-0.081$ o.c. (discussed in more detail in Fig.~\ref{fig:zno2d_traj2}.)}
  \label{fig:zno2d_erm2}
\end{figure}

We now extend our semiclassical ERM analysis to include tunneling from a disc around $\Gamma$ in reciprocal space, as described previously in Sec.~\ref{sec:theory_semiclass}. In our calculations, we choose the radius of the disc for $\ellip=0,0.5,1$ to be $\Delta k=0.15,0.2,0.25$, respectively. In the left (right) panels of Fig.~\ref{fig:zno2d_erm2}, the results for the ERM simulations without (with) the \ehPE{} are shown together with the quantum results in the background (we have made the colorplots transparent to highlight the semiclassical results). For $\ellip=0$ in Figs.~\ref{fig:zno2d_erm2}(a) and \ref{fig:zno2d_erm2}(b), the ERM result is similar to the one in Fig.~\ref{fig:zno2d_cwt}(a): the overall structure is broader, and the short and long type of trajectories are now continuously connected to the higher energy structure. The result with and without the \ehPE{} are also similar, with Fig.~\ref{fig:zno2d_erm2}(b) having some trajectories with higher energy, forming a ``boot'' structure.
We note that in Figs.~\ref{fig:zno2d_cwt}, \ref{fig:zno2d_erm2}, \ref{fig:zno2d_traj2}(a), \ref{fig:hbn2d_anglescan2}(b) and \ref{fig:hbn2d_erm}, we are showing all recolliding trajectories with equal weights. An interesting extension of this work might be to quantify the contribution of individual trajectories by weighting them with their respective tunneling and recollision probabilities.

For $\ellip=0.5$ in Figs.~\ref{fig:zno2d_erm2}(c) and \ref{fig:zno2d_erm2}(d), we clearly observe semiclassical recollisions, in contrast to the case with only the $\Gamma$ point in Fig.~\ref{fig:zno2d_cwt}(b). The half-cycle periodicity and emission times are in overall agreement with the SBE results. To clearly reproduce the triangular structure, it is seen by comparing Figs.~\ref{fig:zno2d_erm2}(c) and \ref{fig:zno2d_erm2}(d) that one has to take into account the \ehPE. It is important to notice that the triangular structure is not due to trajectories tunneled at a single $\vec{k}_0$, but rather trajectories from different $\vec{k}_0$ points that collectively give rise to the full triangular emission structure. This new finding is in stark contrast to HHG in gases and updates the previous conception for HHG in solids where the trajectories tunnelled from the minimum band gap is the only ones that mattered. Interestingly, comparing the $\ellip=0$ case with $\ellip=0.5$, the origin of the triangular structure seems to be due to the class B trajectories.

For the circularly polarized case in Figs.~\ref{fig:zno2d_erm2}(e) and \ref{fig:zno2d_erm2}(f), the six-fold symmetry is clearly reproduced in the semiclassical calculations. Inclusion of the \ehPE{} reproduces the almost vertical structures extending up to \hh 60. For clarity, only the first recollision for each trajectory is counted in Fig.~\ref{fig:zno2d_erm2}; when more recollisions are counted, the ERM reproduces more features in the colorplots (see Fig.~\ref{fig:app_zno2d_erm} in Appendix~\ref{sec:app_suppl}).


\begin{figure}
  \centering
  \includegraphics[width=0.5\textwidth, clip, trim=0 0cm 0 0cm]{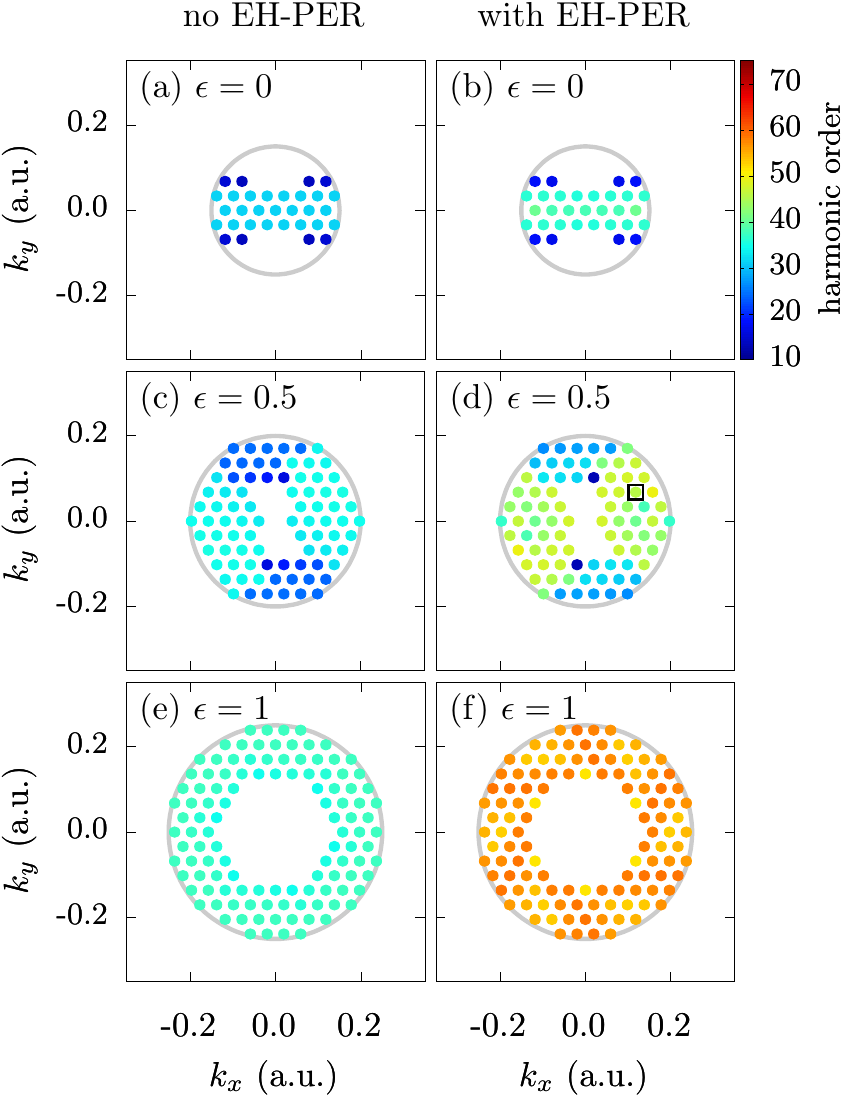}
  \caption{Maximum semiclassical recollision energy for different initial tunneling crystal momenta $\vec{k}_0$ in the BZ. Left (right) panels show the semiclassical results without (with) the inclusion of the \ehPE{} $\vec{F}(t)\cdot\Delta\vec{r}$ in Eq.~\eqref{eq:theory_sfa_3c}, while different row panels correspond to different ellipticities $\ellip$. All $\vec{k}_0$ points contained inside the gray circles are propagated, and missing points indicate no trajectory recollision (or recollision energy $\omega < 11\omega_0$). The $\vec{k}_0$ with the highest recollision energy in (d) is enclosed by a square.}
  \label{fig:zno2d_maxrecol}
\end{figure}

To explore the contributions from different initial tunnel sites $\vec{k}_0$ to the time-frequency profiles and the HHG spectra, we show in Fig.~\ref{fig:zno2d_maxrecol} the maximum recollision energy (colorbar) as a function of the tunnel site in reciprocal space. Each subfigure in Fig.~\ref{fig:zno2d_maxrecol} uses the same ERM data set as the corresponding subfigure in Fig.~\ref{fig:zno2d_erm2}. For example, the ERM calculation in Fig.~\ref{fig:zno2d_erm2}(a) contains all initial $\vec{k}_0$ points inside the gray circle in Fig.~\ref{fig:zno2d_maxrecol}(a) (with radius $\Delta k=0.15$). Missing points inside a gray circle in Fig.~\ref{fig:zno2d_maxrecol} indicate no semiclassical recollisions for that particular $\vec{k}_0$. For the $\ellip=0$ case shown in Figs.~\ref{fig:zno2d_maxrecol}(a) and \ref{fig:zno2d_maxrecol}(b), the recolliding trajectories clearly originate with $\vec{k}_0$ along the $k_x$-axis (laser polarization direction): electron-hole trajectories created too far away from the $k_x$-axis will be driven apart in the $y$-direction in real space and never recollide [see Figs.~\ref{fig:zno2d_struc}(c) and \ref{fig:zno2d_struc}(d)]. For $\ellip=0.5$ in Figs.~\ref{fig:zno2d_maxrecol}(c) and \ref{fig:zno2d_maxrecol}(d), trajectories starting at the $\Gamma$ point clearly do not recollide, in agreement with Fig.~\ref{fig:zno2d_cwt}(b). For the circularly polarized case in Figs.~\ref{fig:zno2d_maxrecol}(e) and \ref{fig:zno2d_maxrecol}(f), only trajectories with $\norm{\vec{k}_0}\gtrsim 0.15$ can recollide, and the six-fold symmetry of the BZ is clearly visible. Note in Fig.~\ref{fig:zno2d_maxrecol} that the larger the ellipticity, the larger the ``hole'' around $\Gamma$ becomes, and the less the trajectories starting near $\Gamma$ contribute to the interband emissions. The maximum recollision energies including the \ehPE{} (right panels in Fig.~\ref{fig:zno2d_maxrecol}) are substantially higher than the calculations without (left panels), in agreement with the results in Fig.~\ref{fig:zno2d_erm2}. Figure~\ref{fig:zno2d_maxrecol} again reinforces our central finding that taking into account only $\vec{k}_0=\Gamma$ is insufficient for the description of HHG in bulk solids with elliptically polarized drivers. While tunneling indeed occurs mostly at $\Gamma$ [e.g. the band gap at $\vec{k}=(0,0,0.1)$ is increased by 13\% compared to $\vec{k}=\Gamma$], the dynamics imposed by the laser and the dispersion relation is such that recollision is prevented.


\begin{figure}
  \centering
  \includegraphics[width=0.49\textwidth, clip, trim=0 0cm 0 0cm]{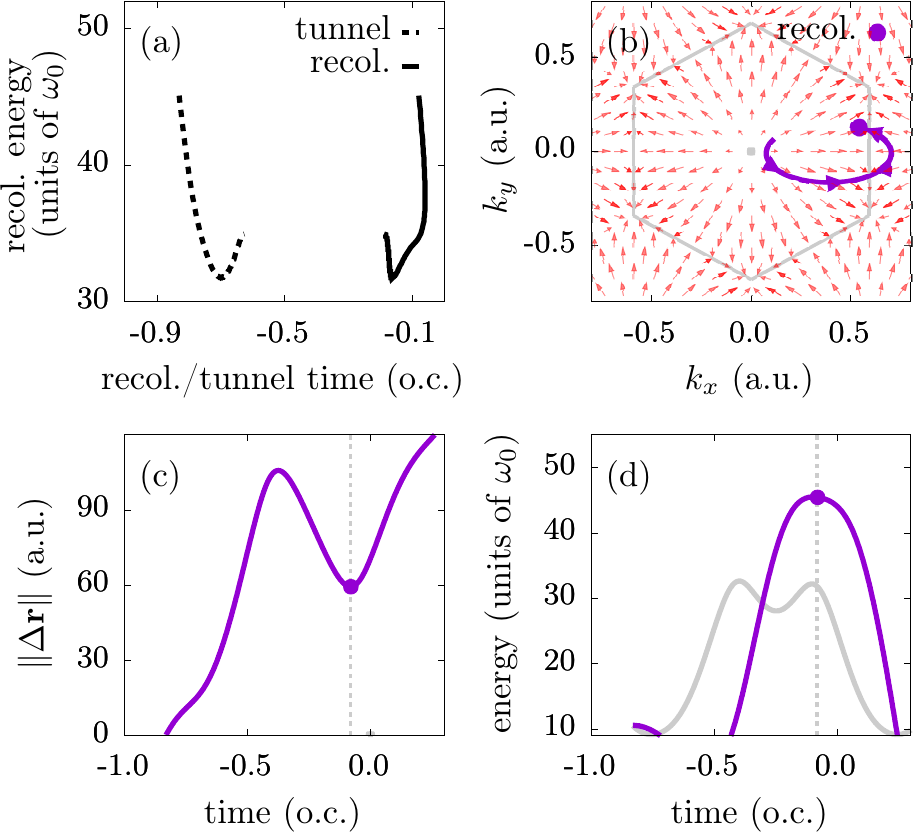}
  \caption{Semiclassical analysis for $\ellip=0.5$ and the trajectories that tunnel at $\vec{k}_0=(0.118, 0.068)$ [enclosed by a square in Fig.~\ref{fig:zno2d_maxrecol}(d)].
    (a) Recollision energy as a function of tunnel and recollision times. (b)-(c) Analysis for the specific trajectory with the highest recollision enegy (tunnel at $s=-0.83$ o.c. and recollide at $t=-0.081$ o.c., red diamonds in Fig.~\ref{fig:zno2d_erm2}): (b) Motion in reciprocal space (purple solid trajectory), and the hole group velocity (red vector field); (c) Electron-hole distance versus time; (d) Energy versus time, with the gray (purple) line showing the energies without (with) the \ehPE.}
  \label{fig:zno2d_traj2}
\end{figure}

For $\ellip=0.5$, the $\vec{k}_0$ with the highest recollision energy is marked with a square in Fig.~\ref{fig:zno2d_maxrecol}(d). Correspondingly, this $\vec{k}_0$ gives rise to the tip of the triangular structure in time-frequency profiles of Fig.~\ref{fig:zno2d_erm2}(d). Figure~\ref{fig:zno2d_traj2}(a) shows the recollision energies versus the tunnel and recollision times for all trajectories originating with this $\vec{k}_0$. Clearly, all resemblances to the short and long trajectories from gas-phase HHG are gone. Instead, the recollision energies versus the recollision times exihibits a highly irregular structure, with harmonics above order $\sim 35$ emitted approximately at the same time.

To give an example, we now focus on the trajectory with the highest recollision energy that tunnels at $s=-0.83$ o.c. and recollides at $t=-0.081$ o.c. (red diamonds in Fig.~\ref{fig:zno2d_erm2}). The time-dependent crystal momentum shown in Fig.~\ref{fig:zno2d_traj2}(b) extends beyond the first BZ, and the electron-hole trajectories are seen to recollide imperfectly with $\norm{\Delta\vec{r}}\approx 60$ in Fig.~\ref{fig:zno2d_traj2}(c). The recollision energy is increased by $\sim 15$ harmonic orders due to the \ehPE{} [Fig.~\ref{fig:zno2d_traj2}(d)], which leads to the correct reproduction of the triangular structure in Fig.~\ref{fig:zno2d_erm2}(d) and not in Fig.~\ref{fig:zno2d_erm2}(c).

To summarize this section, we have shown that for a generic bulk solid with the minimum band gap at $\Gamma$, elliptical drivers enhance the harmonic emissions at high frequencies typically associated with the cut-off region of a harmonic spectrum, and greatly reduce the harmonic intensity in the plateau region. The time-frequency analysis reveals that the highest-order harmonics are not emitted from trajectories tunnelled at $\Gamma$, but rather due to collective emissions originating from many $\vec{k}_0$s near $\Gamma$.

\section{Recollisions in a monolayer bandgap material} \label{sec:hbn2d}
In the previous section, we investigated HHG in a generic model for bulk solids where the minimum band gap is at the high-symmetry point $\Gamma$ of the BZ. In bandgap monolayer matarials, in contrast, the minimum band gap is usually located at the high-symmetry point $K$, with the maximum band gap at $\Gamma$. In this section, we investigate the wavelength and orientation dependence of HHG in a typical topologically trivial monolayer system, using the formalisms presented in Sec.~\ref{sec:theory}.

\subsection{Typical monolayer system: hBN model} \label{sec:hbn2d_model}


\begin{figure}
  \centering
  \includegraphics[width=0.48\textwidth, clip, trim=0 0cm 0 0cm]{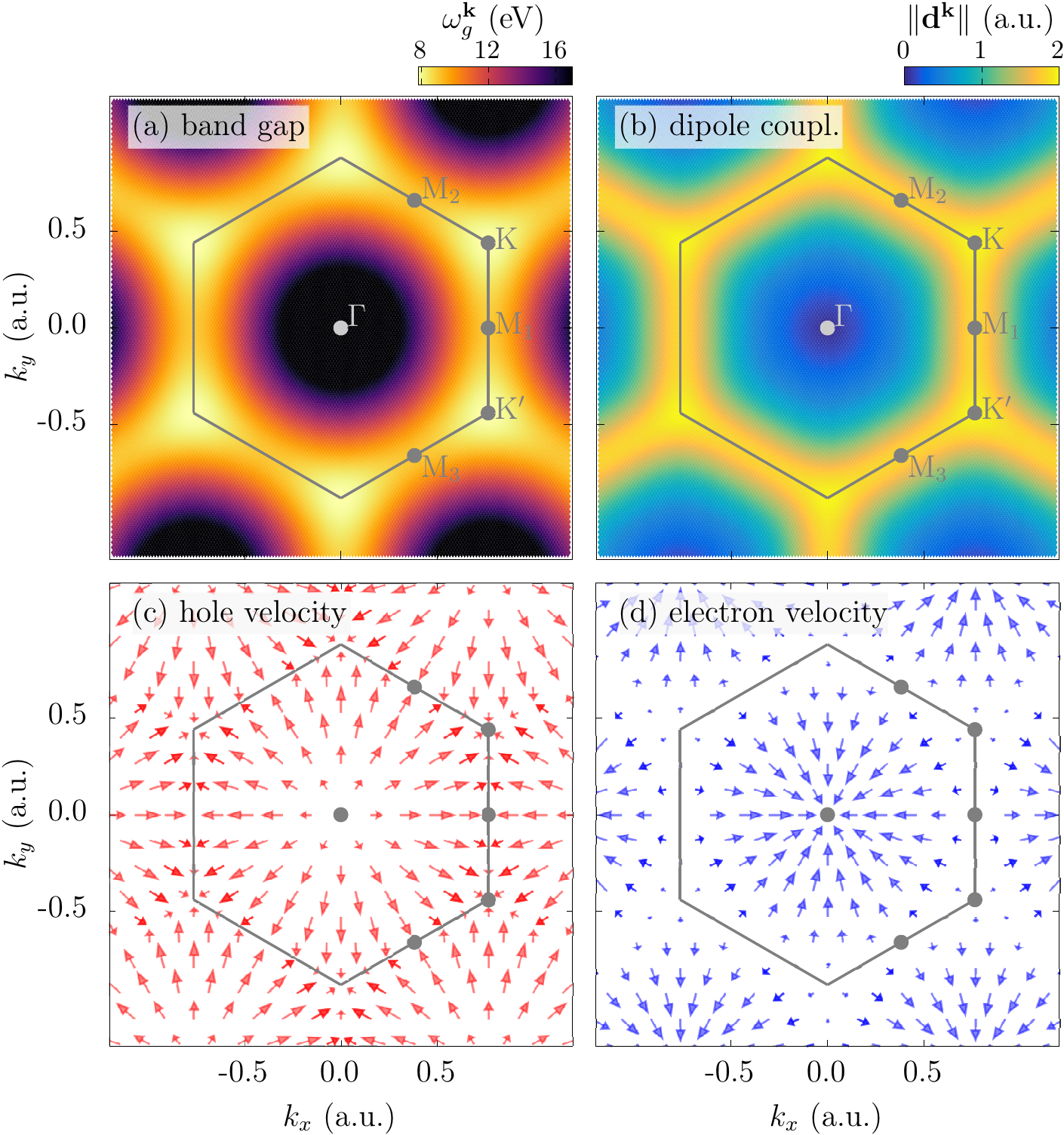}
  \caption{(a) Band gap of monolayer \hBN. (b) Norm of the transition dipole moment. Group velocities of (c) holes in the valence band and (d) of the electrons in the conduction band, without the anomalous velocity. The hexagon in the plots guides the eye and traces the first BZ.}
  \label{fig:hbn2d_struc}
\end{figure}

We use monolayer \hBN{} as an example of a typical monolayer band-gap material. For the band structure calculations, we employ the pseudo potential from \cite{Taghizadeh2017}, and we employ the twisted parallel transport gauge \cite{Vanderbilt2018} to obtain BZ-periodic transition dipole moments and Berry connections \cite{Yue2020pra}. Contrary to the bulk case, the band gap is smallest near the $K$ and $M$ symmetry points and largest at $\Gamma$, as shown in Fig.~\ref{fig:hbn2d_struc}(a). Correspondingly, the norm of the dipole coupling is largest near the $K$ and $M$ points in Fig.~\ref{fig:hbn2d_struc}(b). Figures~\ref{fig:hbn2d_struc}(c) and \ref{fig:hbn2d_struc}(d) show that $\Gamma$ ($K$) acts as a source (sink) and $K$ acts as a sink (source) for the hole (electron) group velocity.

\subsection{Orientation dependence of HHG in \hBN} \label{sec:hbn2d_orientation}


\begin{figure}
  \centering
  \includegraphics[width=0.48\textwidth, clip, trim=0 0cm 0 0cm]{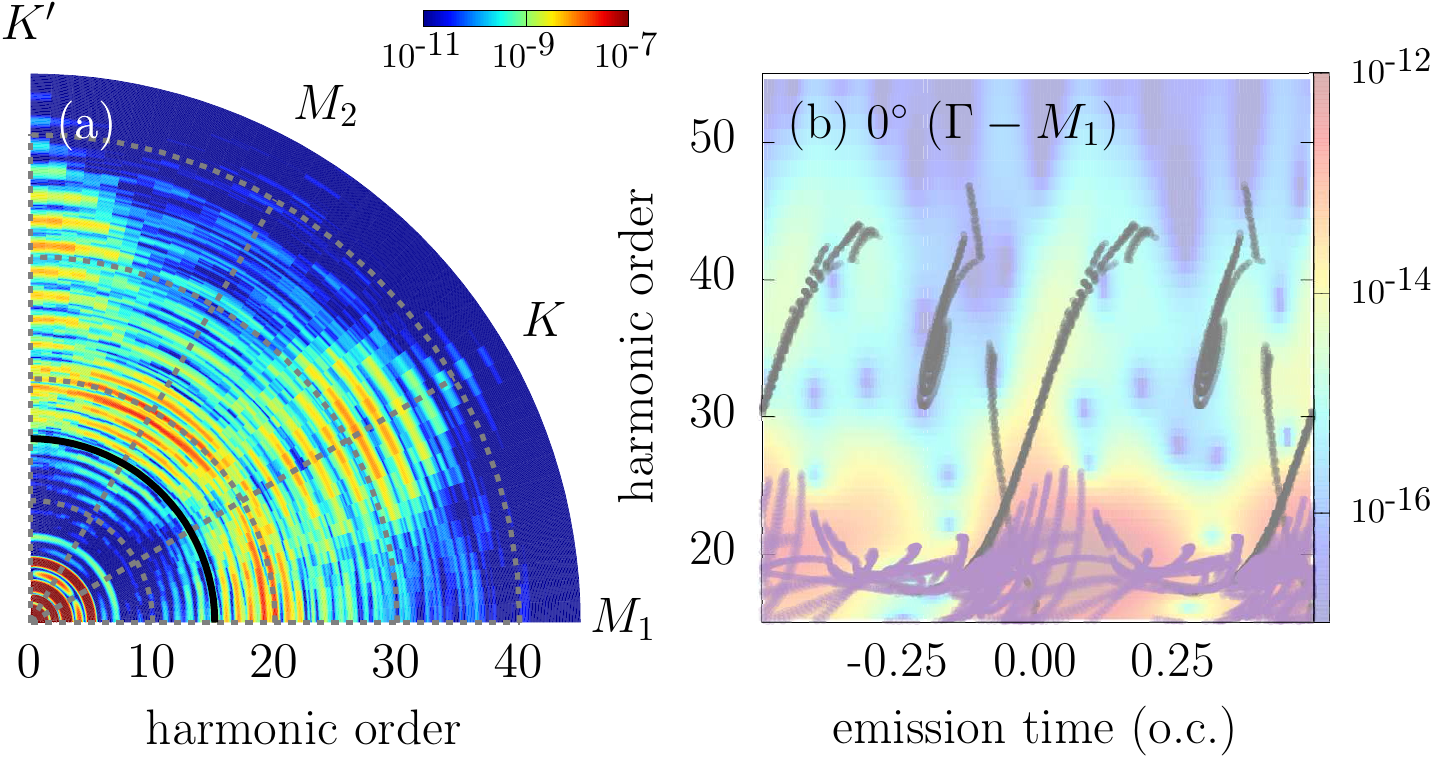}
  \caption{(a) HHG spectrum for the parallel-polarized harmonics as a function of the driver polarization orientation. The pulse parameter of the driver is $\omega_0=0.0190$ ($\lambda=2.4$ $\mu$m), $F_0=0.010$ ($A_0=0.525$), and 5.5 o.c. The black line at $\sim$ \hh 15 indicates the minimum band gap. (a) Time-frequency profile for the $\Gamma-M_1$ driver polarization direction, with the semiclassical ERM result superimposed. For the ERM results, the gray points are recollisions with tunneling around $\vec{k}_0=M_1$ with $\Delta k=0.1$ and the recollision threshold $R_0=15$;  the purple points are recollisions with tunneling around $\vec{k}_0=M_2,M_3,K,K'$.}
  \label{fig:hbn2d_anglescan2}
\end{figure}

We irradiate \hBN{} with linearly polarized infrared pulses and investigate the HHG process with respect to the driver polarization angle $\theta$. The chosen field parameters are $\lambda=2.4$ $\mu$m, $F_0=0.010$ and $\tau$ contains 5.5 o.c. The SBEs are solved with $T_2=10$ fs, and the HHG spectra for the parallel-polarized harmonics are shown in Fig.~\ref{fig:hbn2d_anglescan2}(a). The six-fold symmetry of the BZ is clearly reflected in the spectrum, with stronger yields along the $\Gamma-K$ directions compared to the $\Gamma-M_1$ directions.

Figure~\ref{fig:hbn2d_anglescan2}(b) shows the time-frequency profile for the $\Gamma-M_1$ driver direction with the semiclassical ERM result superimposed \footnote{We mention that for the ERM calculations here, we have neglected the terms in the saddle point equations \eqref{eq:theory_sfa_6} involving $\com^{\vec{k},\pol}$ due to numerical complexities associated with their evaluation. However, due to the agreement between the quantum and semiclassical calculations, as well as the fact that these terms are small in the $\Gamma-K$ case, we believe this is a good approximation}.
Part of the time profile resembles that due to the short trajectories in HHG in gases, with a single ``arm'' extending from \hh 20 to \hh 50 during each half cycle. 
In addition, the most intense part of the radiation is emitted between \hh 15 and \hh 25, at times around $t=0,\pm0.5$ o.c. The gray points show the semiclassical ERM results for trajectories that tunnel around $\vec{k}_0=M_1$ with $\Delta k=0.1$ and $R_0=15$. The single arm in the time-frequency profile is clearly reproduced. Note that the current situation is similar to the atomic HHG case, as well as HHG in bulk solids driven by linearly polarized pulses described in Sec.~\ref{sec:zno2d}. In all these cases, the group velocities [Figs.~\ref{fig:hbn2d_struc}(c) and \ref{fig:hbn2d_struc}(d)] of the trajectories are pointing along the vector potential $\vec{A}(t)$, leading to (almost) perfect recollisions.
The intense features at lower-order harmonics are due to recollisions from trajectories that tunnel near the other symmetry points $\vec{k}_0=M_2,M_3,K,K'$. The ERM results originating from these points are shown in Fig.~\ref{fig:hbn2d_anglescan2}(b) by the purple points and reproduce the intense features very well.
The fact that trajectories originating from $M_1$ lead to much higher recollision energies compared to the other symmetry points can be intuitively predicted by considering the band structure in Fig.~\ref{fig:hbn2d_struc}(a): starting from the $M_1$ symmetry point, the time-dependent crystal momentum $\kappa$ moving along the $\Gamma-M_1$ direction can get closer to the large-band-gap region near the $\Gamma$ point, compared to if one starts from a non-$M_1$ symmetry point.
Our results here show that tunneling from different regions in the BZ can lead to distinct regions in the emission profiles separated in frequency and time. In such cases, the ERM provides a full understanding of the emission dynamics.

The case for $\theta=30^\circ$, i.e. driver polarization along $\Gamma-K$, will be discussed in detail in the next subsection. We here only note that compared to the $\theta=0^\circ$ case discussed here, additional complexities will arise by considering the dispersion relations in Figs.~\ref{fig:hbn2d_struc}(c) and \ref{fig:hbn2d_struc}(d): the group velocities of the electron-hole pairs that start near $\vec{k}_0=M_1$ will no longer be along the vector potential direction ($\Gamma-K$).

\subsection{Wavelength dependence of HHG in \hBN} \label{sec:hbn2d_hhg}


\begin{figure}
  \centering
  \includegraphics[width=0.48\textwidth, clip, trim=0 0cm 0 0cm]{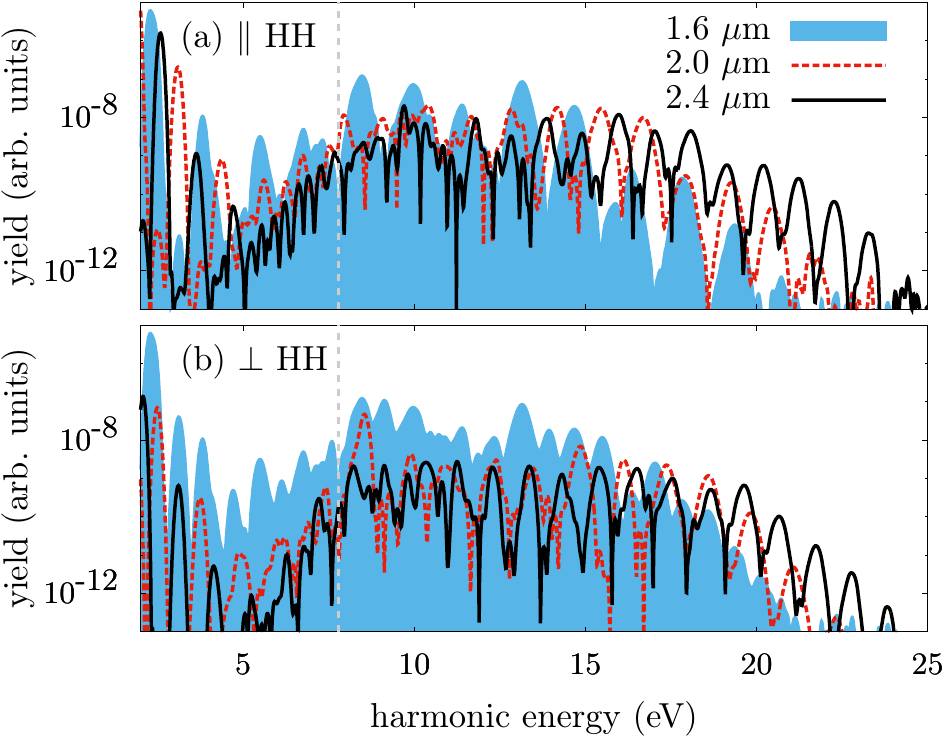}
  \caption{High-harmonic spectra of \hBN{} driven by linearly polarized pulse along $\Gamma$-K, with (a) parallel harmonics and (b) perpendicular harmonics. The field amplitude is fixed at $F_0=0.010$, with the FWHM of the pulse containing 5.5 o.c., and dephasing is set to $T_2=10$ fs.  The vertical dashed lines trace the minimum band-gap energy.}
  \label{fig:hbn2d_wl}
\end{figure}

We irradiate \hBN{} with linearly polarized pulses along the $\Gamma-K$ direction, keeping the field maximum fixed at $F_0=0.010$ and varying the wavelength $\lambda$ from 1.6 $\mu$m to 2.4 $\mu$m, with the FWHM of the pulse $\tau$ chosen to contain 5.5 o.c.. The HHG spectra calculated from the SBEs are shown in Fig.~\ref{fig:hbn2d_wl}(a) and Fig.~\ref{fig:hbn2d_wl}(b) for the parallel- and perpendicular-polarized harmonics, respectively. The HHG spectra extend toward higher harmonic energies with increasing wavelengths, which can be qualitatively understood simply by the larger $A_0$ and consequently the larger excursion of the time-dependent crystal momenta in Eq.~\eqref{eq:theory_sfa_6}.


\begin{figure}
  \centering
  \includegraphics[width=0.48\textwidth, clip, trim=0 0cm 0 0cm]{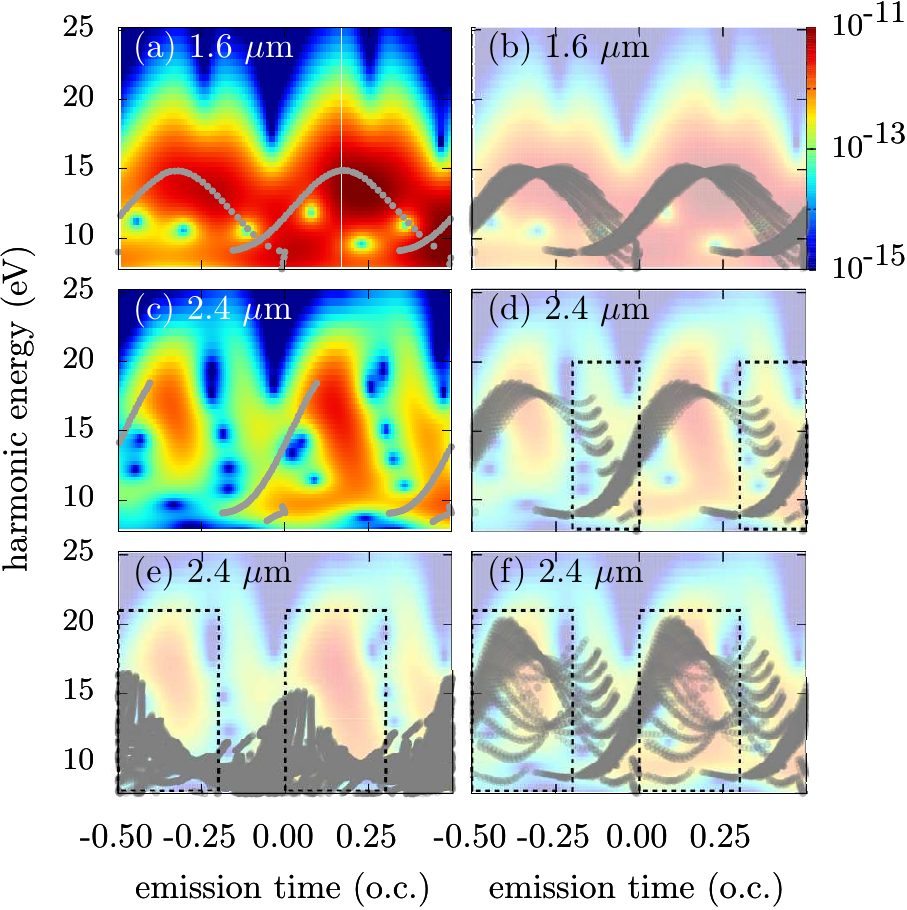}
  \caption{Time-frequency profiles (colormaps) for the parallel-polarized harmonics obtained from the SBE simulations with the semiclassical ERM results superimposed. Panels (a) and (b) are for a $1.6$ $\mu$m driver, while panels (c)-(f) are for a $2.4$ $\mu$m driver. For the ERM results, panels (a) and (c) show the results for tunneling initiated only at $\vec{k}_0=M_1,M_2$, while panels (b) and (d) are for tunneling in discs of radius $\Delta k=0.1$ around the $M_1,M_2$ symmetry points, with the recollision threshold set to $R_0=15$. Panels (e) and (f) show ERM results with the first recollisions and $R_0=30$: (e) results for tunneling from discs around $\vec{k}_0=M_3,K,K'$, while (f) results for tunneling from discs around $\vec{k}_0=M_1,M_2$. The rectangular squares are guides to the eye for the discussions in the text.}
  \label{fig:hbn2d_erm}
\end{figure}

The time-frequency profiles in the colorplots of Fig.~\ref{fig:hbn2d_erm} reveal the emission dynamics of the HHG process. For the 1.6 $\mu$m case in Fig.~\ref{fig:hbn2d_erm}(a) the characteristic double-peak structure during each half-cycle is observed, which was studied in detail in Ref.~\cite{Yue2020PRL}. When the wavelength is increased, during each half-cycle, the double-peak seems to split into two almost-vertical, downwards-sloping structures, as shown in Figs.~\ref{fig:hbn2d_erm}(c). The ERM results in the case of only taking into account the $M_1$ and $M_2$ symmetry points are overlaid on top of the colorplots in Figs.~\ref{fig:hbn2d_erm}(a) and \ref{fig:hbn2d_erm}(c). Clearly, in Fig.~\ref{fig:hbn2d_erm}(a) they are unable to reproduce the double-peak structure and in Fig.~\ref{fig:hbn2d_erm}(c) the semiclassical results seem to be at odds with the colorplot, predicting recollisions at times when there are actually no emissions.

We extend the ERM analysis to include all tunnel points $\vec{k}_0$ in a disc of radius $\Delta k=0.1$ around $M_1$ and $M_2$, shown by the gray dots in Figs.~\ref{fig:hbn2d_erm}(b) and \ref{fig:hbn2d_erm}(d). The recollision threshold is set to $R_0=15$. For the 1600 nm case in Figs.~\ref{fig:hbn2d_erm}(b), it is seen that the double peak structures are attributed to imperfect recollisions for trajectories tunnelled close to the $M$ points \cite{Yue2020PRL}. For the $2.4$ $\mu$m case in Fig.~\ref{fig:hbn2d_erm}(d), we first focus on the downwards-sloping structures enclosed by the rectangular boxes in Fig.~\ref{fig:hbn2d_erm}(d). The ERM results are seen to reproduce these downwards-sloping structures quite well, although they are not continuous, and resemble groups of horizontal lines separated by vertical spacings. These gaps are due to the density of discrete $\vec{k}_0$ points chosen in our simulations: when the density is increased, the empty spacings in the semiclassical results get filled.

We now turn our attention to the prominent downwards-sloping structures in the time profiles that are not reproduced, highlighted by the rectangular boxes in Figs.~\ref{fig:hbn2d_erm}(e) and \ref{fig:hbn2d_erm}(f). We argue that these structures are due to not just contributions from a number of different regions in the BZ, as we have seen above, but also how these contributions interfere with each other. We perform ERM calculations with $R_0=30$ and taking into account the first three recollisions (instead of one). The gray dots in Fig.~\ref{fig:hbn2d_erm}(f) show the ERM results for trajectories with $\vec{k}_0$ near $M_1$ and $M_2$. The prominent downward-sloping structures in the boxes are mostly covered by the ERM results. However, the time-frequency profile contains prominent holes in the left bottom part of the boxes, indicating the absence of harmonic emissions, which are not reproduced by the ERM results. To examine further, we show in Fig.~\ref{fig:hbn2d_erm}(e) the ERM results for recolliding trajectories initiated with $\vec{k}_0$ near the $K$, $K'$ and $M_3$ symmetry points. Recollisions are observed covering the lower left parts of the boxes, exactly in the regions where emissions should be absent according to the quantum results. Since emissions with the same time and harmonic energy should be added coherently, and the trajectories inititated near all the symmetry points ($M_1$, $M_2$, $M_3$, $K$, $K'$) overlap here with widely different phases, they appear to destructively interfere and lead to the absence of emissions. The dominant structures in the time profiles are then reproduced by the parts of the ERM result in Fig.~\ref{fig:hbn2d_erm}(f) that do not overlap with the ERM result in Fig.~\ref{fig:hbn2d_erm}(e). Note that a definite proof of the described destructive interference effect is beyond the ERM and the scope of the current work. Still, the semiclassical method gives us insight on where in the BZ the different trajectories originate, which can lead to, in our opinion, a satisfying understanding of this interference effect and the final dynamics.

Again, in this section we have found that the novel time-frequency profiles for HHG in solids are due to the collective emission of harmonics originating from different $\vec{k}_0$ points in the BZ.

\subsection{Quantum wave packet analysis} \label{sec:hbn2d_wp}



\begin{figure}
  \centering
  \includegraphics[width=0.48\textwidth, clip, trim=0 0cm 0 0cm]{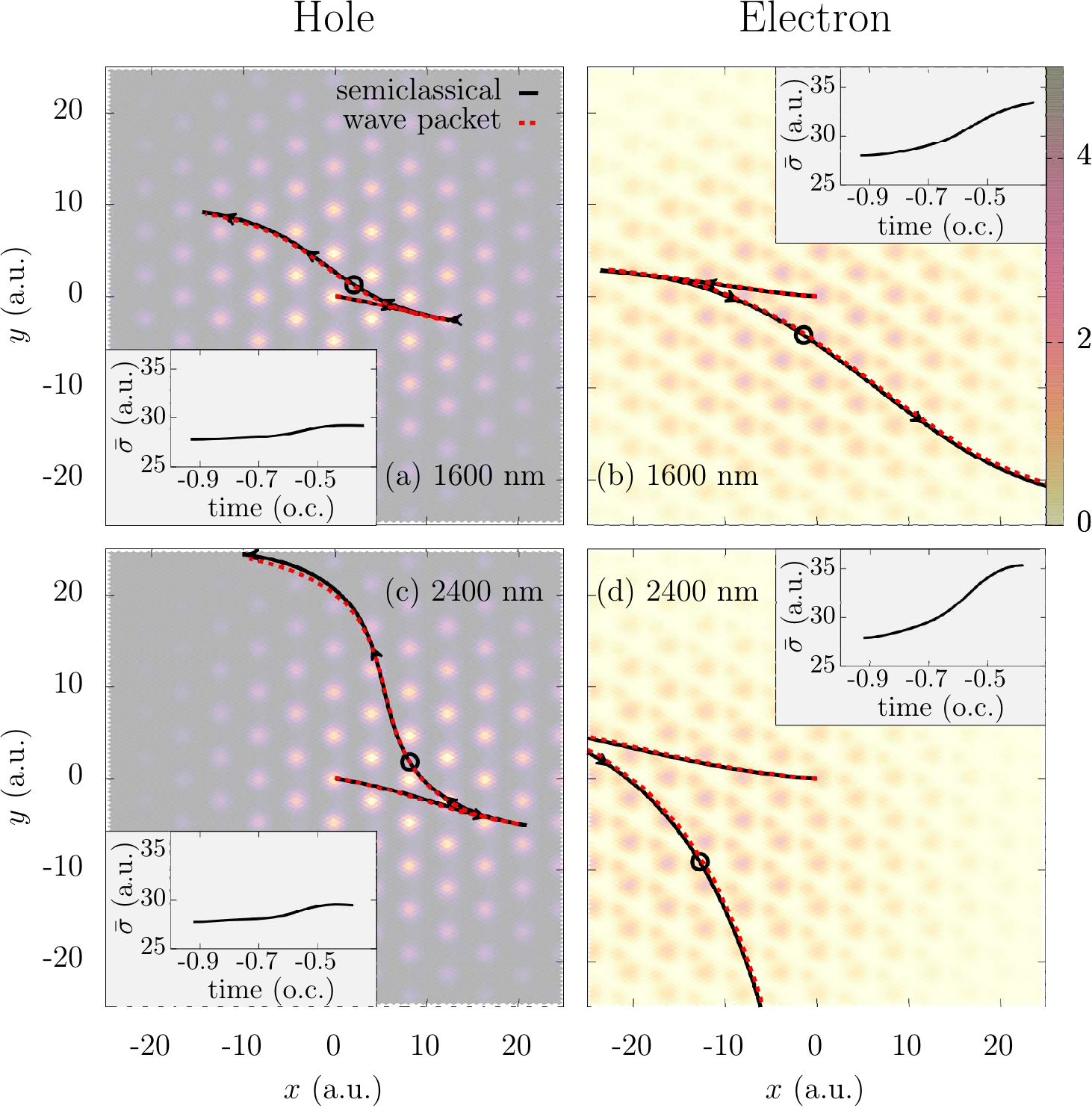}
  \caption{Quantum wave packet results for specific recolliding trajectories that tunnel at $\vec{k}_0=M_1$ with initial FWHM  $0.1$ in reciprocal space. (a),(b) Electron and hole motion for the 1.6 $\mu$m driver pulse and the specific trajectory that tunnels at $s=-0.931$ o.c. and recollides at $t=-0.321$ o.c. The solid black curves show the trajectory real-space motion obtained using ERM, while the dashed red curve show the expectation value $\left<\vec{r}\right>$ of the wave packets. The real-space position at the time of recollision $t$ is indicated by the circle, with the colorplot showing the wave packet probability density at $t$. The insets show the widths of the wave packets from $s$ to $t$. (c),(d) Same as (a) and (b), but for the 2.4 $\mu$m driver and the trajectory that tunnels at $s=-0.922$ o.c. and recollides at $t=-0.3634$ o.c.}
  \label{fig:hbn2d_wp}
\end{figure}

We have shown that the semiclassical ERM model is able to capture the emission dynamics of the HHG process in solids. The imperfect recollision and the origin of the \ehPE{} can be interpreted in the context of spatially extended wave packets at the time of recollision.
We employ the formalism described in Sec.~\ref{sec:theory_wp} to construct and visualize such wave packets. We assume tunneling at $\vec{k}_0=M_1$ with initial FWHM $0.1$ in reciprocal space (approximated by Zener tunneling). For the 1.6 $\mu$m driver polarized along $\Gamma-K$, we consider the trajectory with the highest recollision energy which tunnels at $s=-0.931$ o.c. and recollides at $t=-0.321$ o.c. The dashed red curves in Figs.~\ref{fig:hbn2d_wp}(a) and \ref{fig:hbn2d_wp}(b) show the real-space motion $\left<\vec{r}\right>$ of the hole and electron wave packets, respectively. The quantum wave packet results are seen to agree perfectly with the semiclassical ERM results shown by the solid black curves. The hollow circles indicate the real-space position at the time of recollision $t$, and the colorplots show the wave packet probability density at $t$. The wave packets have a large width and extend over many lattice sites. The time dependence of the wave packet width $\bar{\sigma}$ from $s$ to $t$ is shown in the insets of Figs.~\ref{fig:hbn2d_wp}(a) and \ref{fig:hbn2d_wp}(b). Due to the lower effective mass of the conduction band, the electron wave packet moves a greater distance compared to the hole and spreads more. Note that at the time of recollision, the electron and hole wave packets clearly occupy the same spatial region and overlap.

Figures \ref{fig:hbn2d_wp}(c) and \ref{fig:hbn2d_wp}(d) show the wave packet results for the 2.4 $\mu$m driver and the trajectory that attains the highest recollision energy. The quantum wave packet and the semiclassical motion are again in full agreement. Due to the longer half-cycle and larger $A_0$ compared to the 1.6 $\mu$m case, the electron and hole are driven apart further along the $y$-direction according to the group velocities in Fig.~\ref{fig:hbn2d_struc} before the vector potential changes sign, leading to a larger recollision distance. The large electron-hole spatial separation at the time of recollision, however, does not prevent their spatial overlap, as evidenced by the the wave packet densities. Due to longer time duration between tunneling and recollision for the 2.4 $\mu$m, the wave packets spread more compared to the 1.6 $\mu$m cases (insets of Fig.~\ref{fig:hbn2d_wp}).


\section{Conclusion and outlook} \label{sec:hbn2d_conclusion}
We have presented a recollision formalism for HHG in solids that conclusively shows that in many realistic situations the harmonic spectrum and emissions are not due to tunneling at the minimum band gap, but instead due to the collective effect of trajectories originating near different symmetry points in the BZ. Indeed, for the example of HHG in a bulk solid with elliptical drivers, we showed that the electron-hole pairs created at $\Gamma$ do not recollide at all and contribute nothing to the highest-order harmonics. For monolayer materials with hexagonal symmetry, the highest order harmonics originate not from the minimum band gap at the $K$ symmetry points, but near the $M$ points. In addition, we found that the HHG for different driver orientations results in very distinct time-frequency profiles, and we showed that this is due to collective emissions from many different reciprocal-space tunneling sites. 
Interestingly, for certain driver orientations, different parts of the emission profiles can be ascribed to electrons initially tunneling near different symmetry points in the BZ, allowing for future prospects of probing the BZ tunneling regions.
We also showed that the imperfect recollisions leading to the electron-hole polarization energies are important for the correct description of the harmonic emissions, a result which is further supported by our quantum wave packet constructions.
Generally, imperfect recollisions and \ehPE{} will be important whenever the electron-hole separation vector $\Delta \vec{r}$ [Eq.~\eqref{eq:theory_sfa_4a}] is nonzero (the different Cartesian components of $\Delta \vec{r}$ could be zero at different times), i.e. whenever the direction of motion of the time-dependent crystal momentum
\begin{equation}
  \label{eq:concl_1}
  \partial_{t'} \vec{\kappa}(t,t')=-\vec{F}(t')
\end{equation}
is not along the instantaneous group velocities $ \vec{v}_n^{\vec{\kappa}(t, t')}$. 
Thus, situations where solid-state HHG in topogolically-trivial systems differ significantly from gas-phase HHG can be summerized by two simple rules of thumb: (A) when the instantaneous carrier group velocities are not along the electric-field polarization direction; (B) when the time-dependent crystal momentum goes beyond the BZ boundaries and induces Bragg reflections.

Our work illustrates the complexity of HHG in solids compared to HHG in the gas phase and broadens the notion of which parts of the BZ contribute to the emission process - in particular that the most important symmetry point is not always at the minimum band gap. The detailed knowledge gained from the collective emissions responsible for the novel time-frequency profiles, aside from the fundamental perspective, will have impact on future experiments that involves phase matching and ultrafast spectroscopy. The strong interest in the generation of elliptically-polarized harmonics will also benefit from this work. Furthermore, the fact that tunneling from different regions in the BZ leads to distinct harmonic time-frequency characteristics can potentially facilitate the all-optical reconstruction of the band structure not only near the minimum band gap as demonstrated in \cite{Vampa2015PRL}, but near all relevant symmetry points in the BZ.


\acknowledgements
The authors acknowledge support from the National Science Foundation, under Grant No. PHY1713671
and useful interactions with Guilmot Ernotte.
Portions of this research were conducted with high performance computational resources provided by the Louisiana Optical Network Infrastructure (http://www.loni.org).

\appendix

\section{Derivations} \label{sec:app_derivs}

In this appendix, we provide some more details on some of the derivation steps in Sec.~\ref{sec:theory}.

\subsection{Saddle-point equations} \label{sec:app_derivs_1}

In the two-band approximation, the SBEs in Eq.~\eqref{eq:theory_sbe_1} reduces to
\begin{subequations}
  \label{eq:app_saddle_1}
  \begin{align}
    \dot{\rho}_{vv}^{\vec{K}} = & i \vec{F}\cdot
                                  \vec{d}^{\vec{K}+\vec{A}}\rho_{vc}^{\vec{K}} + \text{c.c.}\\
    \dot{\rho}_{cc}^{\vec{K}} = & -i \vec{F}\cdot
                                  \vec{d}^{\vec{K}+\vec{A}}\rho_{vc}^{\vec{K}} + \text{c.c.}\\
    \dot{\rho}_{cv}^{\vec{K}} = & \left[
                                  - i \omega_g^{\vec{K}+\vec{A}}
                                  - i \vec{F} \cdot  \Delta \vec{\berryc}^{\vec{K}+\vec{A}}
                                  - T_2^{-1}
                                  \right] \rho_{cv}^{\vec{K}} \nonumber \\
                                &- i \left(\rho_{vv}^{\vec{K}}-\rho_{cc}^{\vec{K}}\right) \vec{F} \cdot \vec{d}^{\vec{K}+\vec{A}},
  \end{align}
\end{subequations}
where $\Delta\berryc^{\vec{k}}\equiv\berryc_c^{\vec{k}}-\berryc_v^{\vec{k}} $ and for notational convenience the explicit time-dependencies in $\vec{A}(t)$, $\vec{F}(t)$ and $\rho_{mn}^{\vec{K}}(t)$ have been omitted. Now we make the approximation of minimum population transfer for the conduction band, $\rho_{vv}^{\vec{K}}-\rho_{cc}^{\vec{K}}\approx 1$, the formal solutions to Eq.~\eqref{eq:theory_sbe_1} read
\begin{subequations}
  \label{eq:app_saddle_2}
  \begin{align}
    \rho_{vv}^{\vec{K}}(t) = & i \int^t ds \vec{F}(s) \cdot \vec{d}^{\vec{K}+\vec{A}(s)}\rho_{vc}^{\vec{K}}(s) + \text{c.c.}\\
    \rho_{cc}^{\vec{K}}(t) = & -i \int^t ds \vec{F}(s) \cdot \vec{d}^{\vec{K}+\vec{A}(s)}\rho_{vc}^{\vec{K}}(s) + \text{c.c.}\\
    \rho_{cv}^{\vec{K}}(t) = & -i \int^t ds
                               \vec{F}(s) \cdot \vec{d}^{\vec{K}+\vec{A}(s)} e^{- T_2^{-1} (t-s)}\\
                             & \times e^{- i\int_s^t
                               \left[ \omega_g^{\vec{K}+\vec{A}(t')} + \vec{F}(t') \cdot \Delta\vec{\berryc}^{\vec{K}+\vec{A}(t')} \right]
                               dt'}.
  \end{align}
\end{subequations}

The expression for the interband current in Eq.~\eqref{eq:theory_sfa_1} is then obtained by inserting the solution \eqref{eq:app_saddle_2} into Eq.~\eqref{eq:theory_sbe_2a}, and transforming into the fixed frame $\vec{k}=\vec{K}+\vec{A}(t)$.

The saddle point conditions for the interband harmonics are obtained by taking the partial derivatives of $S^\pol(\vec{k},t,s)-\omega t$. The derivative with respect to $s$ reads (henceforth the dephasing time $T_2$ is ignored)
\begin{equation}
  \label{eq:app_saddle_3}
  \begin{aligned}
    \partial_{s} &\left[ S^\pol(\vec{k}, t, s) - \omega t \right] \\
    = &-\omega_g^{\vec{\kappa}(t, s)} - \vec{F}(s) \cdot \Delta\vec{\berryc}^{\vec{\kappa}(t, s)}
    - \partial_s \beta^{\vec{\kappa}(t, s)} \\
    = &-\omega_g^{\vec{\kappa}(t, s)} - \vec{F}(s) \cdot \vec{\ccom}^{\vec{\kappa}(t,s)},
  \end{aligned}
\end{equation}
where we used $\partial_s\beta^{\vec{\kappa}(t,s)}=-\vec{F}(s)\cdot\nabla_{\vec{k}}\beta^{\vec{k}}|_{\vec{k}=\vec{\kappa}(t,s)}$, and $\vec{\ccom}^{\vec{\kappa}(t,s)}$ is defined in Eq.~\eqref{eq:theory_sfa_5b}.
The derivative with respect to $\vec{k}$ reads
\begin{equation}
  \label{eq:app_saddle_4}
  \begin{aligned}
    \nabla_{\vec{k}} & \left[ S^\pol(\vec{k}, t, s) - \omega t \right] \\
    = & \int_s^t
    \left\{ \nabla_{\vec{k}}\omega_g^{\vec{\kappa}(t, t')}
      + \nabla_{\vec{k}} \left[ \vec{F}(t') \cdot \Delta\vec{\berryc}^{\vec{\kappa}(t,t')} \right] \right\}  dt' \\
    & + \nabla_{\vec{k}} \alpha^{\vec{k},\pol} - \nabla_{\vec{k}} \beta^{\vec{\kappa}(t, s)} \\
    = & \int_s^t
    \left\{
      \vec{v}_c^{\vec{\kappa}(t, t')} - \vec{v}_v^{\vec{\kappa}(t, t')}
    \right\}
    dt'
    - \left. \Delta\vec{\berryc}^{\vec{\kappa}(t,t')} \right|_{t'=s}^{t'=t} \\
    & + \nabla_{\vec{k}} \alpha^{\vec{k},\pol} - \nabla_{\vec{k}} \beta^{\vec{\kappa}(t, s)} \\
    = & \Delta\vec{r} - \vec{\com}^{\vec{k},\mu} + \vec{\ccom}^{\vec{\kappa}(t,s)},
  \end{aligned}
\end{equation}
with $\vec{\com}^{\vec{k},\pol}$, $\Delta \vec{r}$ and $\vec{v}_n^{\vec{\kappa}(t,t')}$ defined in Eqs.~\eqref{eq:theory_sfa_3} and \eqref{eq:theory_sfa_4}, and in the second equality we used the identity $\nabla(\vec{A}\cdot \vec{B})=(\vec{A}\cdot \nabla)\vec{B} + (\vec{B} \cdot \nabla)\vec{A} + \vec{A} \times (\nabla \times \vec{B}) + \vec{B} \times (\nabla \times \vec{A})$, such that
\begin{equation}
  \label{eq:app_saddle_5}
  \begin{aligned}
    &\int_s^t \nabla_{\vec{k}}\left[ \vec{F}(t') \cdot \vec{\berryc}_{n}^{\vec{\kappa}(t,t')} \right] dt' \\
    & = \int_s^t \left\{ \left[ \vec{F}(t') \cdot \nabla_{\vec{k}} \right] \vec{\berryc}_n^{\vec{\kappa}(t, t')}
      + \vec{F}(t') \times \left[ \nabla_{\vec{k}} \times \vec{\berryc}_n^{\vec{\kappa}(t,t')} \right] \right\} dt' \\
    & = \int_s^t \left[ -\partial_{t'} \vec{\berryc}_n^{\vec{\kappa}(t, t')} + \vec{F}(t') \times \vec{\Omega}_n^{\vec{\kappa}(t, t')}\right] dt' \\
    & = -\left. \vec{\berryc}_n^{\vec{\kappa}(t, t')} \right|_{t'=s}^{t'=t} + \int_s^t \left[ \vec{F}(t') \times \vec{\Omega}_n^{\vec{\kappa}(t, t')} \right] dt'.
  \end{aligned}
\end{equation}
The derivative with respect to $t$ reads
\begin{equation}
  \label{eq:app_saddle_6}
  \begin{aligned}
    \partial_{t} & \left[S^\pol(\vec{k}, t, s) - \omega t \right] \\
    = & \omega_g^{\vec{k}}
    + \vec{F}(t) \cdot \Delta\vec{\berryc}^{\vec{k}}
    - \nabla_{\vec{\kappa}} \beta^{\vec{\kappa}(t,s)} \cdot \vec{F}(t) - \omega \\
    & +  \int_s^t
    \partial_t\left[ \omega_g^{\vec{\kappa}(t, t')} + \vec{F}(t') \cdot \Delta \vec{\berryc}^{\vec{\kappa}(t, t')}  \right]
    dt' \\
    = & \omega_g^{\vec{k}}
    + \vec{F}(t) \cdot  \left[ \Delta \vec{\berryc}^{\vec{k}} - \nabla_{\vec{\kappa}} \beta^{\vec{\kappa}(t, s)} \right]
    - \omega \\
    & +  \int_s^t \nabla_{\vec{\kappa}}
    \left[ \omega_g^{\vec{\kappa}(t, t')} + \vec{F}(t') \cdot \Delta \vec{\berryc}^{\vec{\kappa}(t, t')}  \right]
    dt' \cdot \vec{F}(t) \\
    =  & \omega_g^{\vec{k}}
    + \vec{F}(t) \cdot \left[ \Delta \vec{\berryc}^{\vec{k}} - \nabla_{\vec{\kappa}} \beta^{\vec{\kappa}(t, s)} \right]
    - \omega \\
    & + \int_s^t \nabla_{\vec{\kappa}} \omega_g^{\vec{\kappa}(t, t')} dt' \cdot \vec{F}(t)
    - \left.  \Delta \vec{\berryc}^{\vec{\kappa}(t, t')} \right|_{t'=s}^{t'=t} \cdot \vec{F}(t) \\
    & + \int_s^t \left[ \vec{F}(t') \times \left( \vec{\Omega}_c^{\vec{\kappa}(t, t')} - \vec{\Omega}_v^{\vec{\kappa}(t, t')} \right) \right]
    dt' \cdot \vec{F}(t) \\
    = & \omega_g^{\vec{k}}
    + \vec{F}(t) \cdot \left( \Delta \vec{\berryc}^{\vec{\kappa}(t,s)} - \nabla_{\vec{\kappa}} \beta^{\vec{\kappa}(t, s)} \right)
    - \omega \\
    & + \int_s^t \left( \vec{v}_2^{\vec{\kappa}(t,t')} - \vec{v}_1^{\vec{\kappa}(t,t')}  \right) dt' \cdot \vec{F}(t) \\
    = & \omega_g^{\vec{k}}
    + \vec{F}(t) \cdot \left[ \ccom^{\vec{\kappa}(t, s)} + \Delta\vec{r}  \right]
    - \omega.
  \end{aligned}
\end{equation}
The saddle-point conditions in Eq.~\eqref{eq:theory_sfa_3} are then obtained by setting Eqs.~\eqref{eq:app_saddle_3}, \eqref{eq:app_saddle_4} and \eqref{eq:app_saddle_6} to zero.

\subsection{Structure-gauge invariance of $\com^{\vec{k},\pol}$ and $\ccom^{\vec{k}}$} \label{sec:app_derivs_2}

Under the gauge transformation $\ket{u_n^{\vec{k}}} \rightarrow \ket{u_n^{\vec{k}}}e^{i\varphi_n^\vec{k}}$, with $n=v,c$, the relevant quantities transform as
\begin{subequations}
  \begin{align}
    \berryc_n^{\vec{k}} & \rightarrow \berryc_n^{\vec{k}} - \nabla_{\vec{k}}\varphi_n^{\vec{k}} \\
  \vec{d}^{\vec{k}} & \rightarrow \vec{d}^{\vec{k}} e^{-i(\varphi_c^{\vec{k}} - \varphi_v^{\vec{k}})} \\
  \alpha^{\vec{k},\pol} 
                      & \rightarrow \alpha^{\vec{k},\pol} - \varphi_c^{\vec{k}} + \varphi_v^{\vec{k}} \\
  \beta^{\vec{k}} 
                      & \rightarrow \beta^{\vec{k}} - \varphi_c^{\vec{k}} + \varphi_v^{\vec{k}} \\
  \com^{\vec{k},\pol}   & \rightarrow 
  \com^{\vec{k},\pol} \label{eq:app_gauge_1e} \\ 
  \ccom^{\vec{k}} & \rightarrow  
      \ccom^{\vec{k}} \label{eq:app_gauge_1f},
  \end{align}
\end{subequations}
where the transforms in Eqs.~\eqref{eq:app_gauge_1e} and \eqref{eq:app_gauge_1f} are obtained by using the definitions in Eq.~\eqref{eq:theory_sfa_5}.

\subsection{Approximation of tunneling width in WPT} \label{sec:app_derivs_3}



The Landau-Zener tunneling probability \cite{Kane1960, Gauthey1997, Wu2016} reads
\begin{equation}
  \label{eq:app_tun_1}
  P^{\vec{k}} \propto \exp \left[-\frac{\pi \omega_g^{\vec{k}}}{4\abs{\vec{E}\cdot\vec{d}^{\vec{k}}} } \right],
\end{equation}
with $\omega_0$ the laser carrier frequency and $\vec{E}$ chosen at a time when $\abs{\vec{E}\cdot\vec{d}^{\vec{k}}}$ is maximal. In our WPT calculations, we start with a Gaussian wave packet in reciprocal space, with the FWHM width approximated by the FWHM of the above formula.

\subsection{Evaluation of the real-space wave packet} \label{sec:app_derivs_4}
We show here more details on our evaluation of the real-space wave packets. Insertion of the Houston state \eqref{eq:theory_wp_2} into the expression for the wave packet in Eq.~\eqref{eq:theory_wp_1} yields
\begin{equation}
  \label{eq:app_wp_1}
  \Psi_e(\vec{r}, t) = \sum_{\vec{K}\in BZ} a_e^{\vec{K}}(t) u_c^{\vec{K}-q\vec{A}(t)}(\vec{r}) e^{i\vec{K}\cdot\vec{r}},
\end{equation}
which is seen to not be on the form of a Fourier transform, making it expensive for numerical evaluations. 

Often, the $u_c^{\vec{K}}$ functions are given in the Fourier basis (as is the case for our \hBN{} calculations),
\begin{equation}
  \label{eq:app_wp_2}
  u_c^{\vec{K}}(\vec{r}) = \sum_{\vec{G}}u_{c\vec{G}}^{\vec{K}}e^{i\vec{G}\cdot\vec{r}},
\end{equation}
with the sum running over the reciprocal lattice vectors $\vec{G}$. In the twisted parallel transport gauge, the Fourier coefficients satisfy
\begin{equation}
  \label{eq:app_wp_3}
  \begin{aligned}
    \ket{u_c^{\vec{K}+\vec{b}_i}}
    & = e^{-i\vec{b}_i\cdot\vec{r}}\ket{u_c^{\vec{K}}} \\
    \Leftrightarrow
    u_{c\vec{G}}^{\vec{K}+\vec{b}_i}
    & =\bra{e^{i\vec{G}\cdot\vec{r}}} e^{-i\vec{b}_i\cdot\vec{r}}\ket{u_c^{\vec{K}}} \\
    & =\sum_{\vec{G}'} V_{\text{cell}}^{-1}\int_{\text{cell}}e^{i(-\vec{G}-\vec{b}_i+\vec{G}')\cdot\vec{r}} d\vec{r} u_{c\vec{G}'}^{\vec{K}} \\
    & = \sum_{\vec{G}'} \delta_{\vec{G}',\vec{G}+\vec{b}_i} u_{c\vec{G}'}^{\vec{K}} \\
    & = u_{c,\vec{G}+\vec{b}_i}^{\vec{K}}.
  \end{aligned}
\end{equation}
From the above, we have $u_{c\vec{0}}^{\vec{K}+\vec{G}} = u_{c\vec{G}}^{\vec{K}}$, and the wave packet expression can be rewritten
\begin{equation}
  \label{eq:app_wp_4}
  \Psi_e(\vec{r}, t) = \sum_{\vec{K}'\in \text{lattice}} a_e^{\vec{K}'}(t) u_{c\vec{0}}^{\vec{K}'-q\vec{A}(t)}(\vec{r}) e^{i\vec{K}'\cdot\vec{r}},
\end{equation}
where $\vec{K}'\equiv \vec{K}+\vec{G}$ now runs over the entire reciprocal lattice and $a_e^{\vec{K}'}(t) = a_e^{\vec{K}}(t)$.

The crystal momenta $\vec{K}'$ and real-space coordinates $\vec{r}$ are in our calculations given in the basis of the reciprocal and real-space lattice vectors, respectively, such that
\begin{equation}
  \label{eq:app_wp_5}
  \vec{K}'\cdot\vec{r}
  = \left(\sum_{d=1}^DK_d'\uv{b}_d \right) \cdot \left(\sum_{d=1}^Dr_d\uv{a}_d \right) 
  = \sum_{d=1}^{D}K_d'\eta_d,
\end{equation}
where we have defined $\eta_d\equiv r_d 2 \pi/(\abs{\vec{a}_d}\abs{\vec{b}_d})$.
Writing the sum in Eq.~\eqref{eq:app_wp_4} as an integral and taking into account the Jacobian $\vec{J}_{\vec{K}'}$ of the coordinate transformation for $(K'_x,K'_y,K'_z) \rightarrow (K_1',K_2',K_3')$, we can write the wave packet as
\begin{equation}
  \label{eq:app_wp_6}
  \begin{aligned}
    \Psi_e(\vec{\eta}, t) = \det(\vec{J}_{\vec{K}'})\int d\vec{K}' a_e^{\vec{K}'}(t) u_{c\vec{0}}^{\vec{K}'-q\vec{A}(t)}(\vec{\eta}) e^{i\sum_{d=1}^DK_d' \eta_d},
  \end{aligned}
\end{equation}
which we recognize as a multi-dimensional Fourier transform that can be treated using the standard fast-Fourier-transform algorithms.


\section{Supplemental calculations} \label{sec:app_suppl}
\begin{figure}
  \centering
  \includegraphics[width=0.5\textwidth, clip, trim=0 0cm 0 0cm]{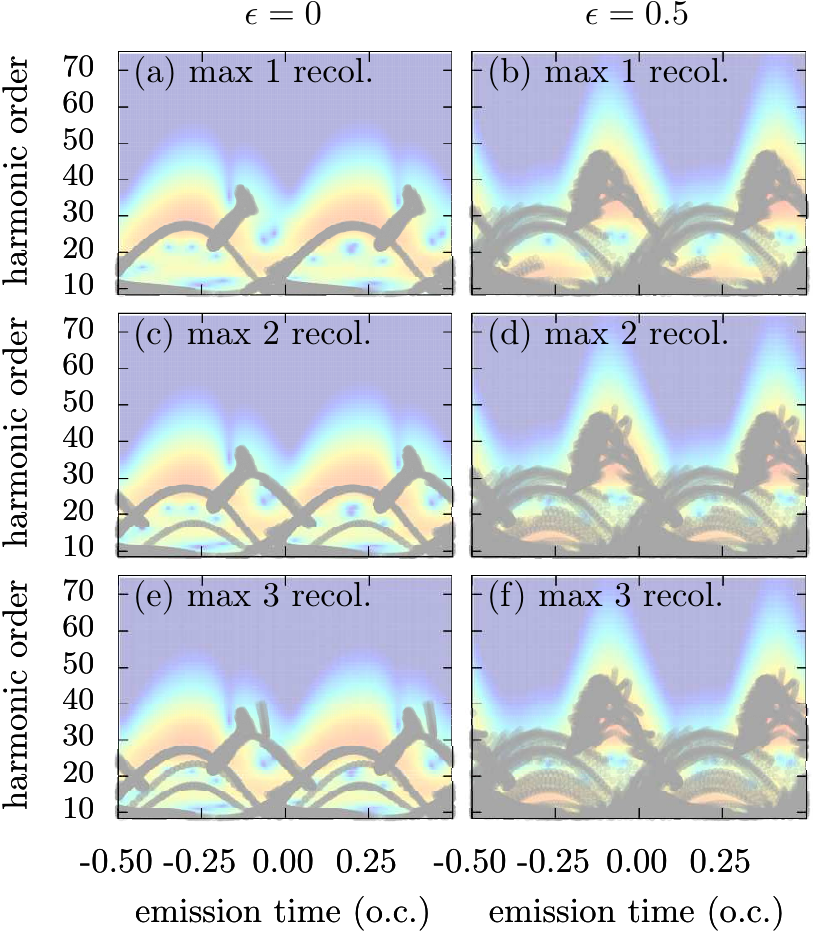}
  \caption{Semiclassical recollisions energies versus recollision times obtained with the ERM (gray dots). Left (right) panels are for $\ellip=0$ ($\ellip=0.5$), while different row panels correspond to different number of allowed recollisions in the calculations.
    The background show the time-frequency profiles from Fig.~\eqref{fig:zno2d_cwt} using the same color scale.}
  \label{fig:app_zno2d_erm}
\end{figure}

To discuss the role of multiple recollisions, we show in Fig.~\ref{fig:app_zno2d_erm} the semiclassical results plotted on top of the time-frequency profiles obtained from the SBEs, for maximum one, two and three recollisions. For both $\ellip=0$ (left panels) and $\ellip=0.5$ (right panels), more features are reproduced in the case of maximum two recollisions compared to maximum one recollision. The case of maximum three recollisions include features not seen in the time profiles, which is understandable due to the low probability of these events.


\bibliographystyle{apsrev4-1}

%

\end{document}